\documentclass[a4paper,fleqn,usenatbib]{mnras}

\usepackage[T1]{fontenc}
\usepackage{ae,aecompl}
\usepackage{graphicx}	
\usepackage{amsmath}	
\usepackage{amssymb}	
\usepackage{multirow}
\title[IC 711 -- the longest head-tail radio galaxy]
{GMRT observations of IC~711 -- the longest head-tail radio galaxy known}
\author[S. Srivastava \& A. K. Singal]{Shweta Srivastava\thanks{E-mail: srivashweta@gmail.com} and Ashok K. Singal\thanks{E-mail:  ashokkumar.singal@gmail.com}\\
{Astronomy and Astrophysics Division, Physical Research Laboratory, 
Navrangpura, Ahmedabad - 380 009, India}\\
}

\date{Accepted XXX. Received YYY; in original form ZZZ}

\pubyear{2018}

\begin{document}
\label{firstpage}
\pagerange{\pageref{firstpage}--\pageref{lastpage}}
\maketitle

\begin{abstract} 
We present low-frequency, GMRT observations at 240, 610 and 1300 MHz of IC~711, a narrow angle tail (NAT) radio galaxy. The total angular extent of the radio emission, $\sim 22$ arcmin, corresponds to a projected linear size of $\sim 900$ kpc, making it the longest among the known head-tail radio galaxies.
The objectives of the GMRT observations were to investigate the radio morphology, especially of the long tail structure, at low frequencies. 
The radio structure, especially initial $\sim 10$ arcmin of tail being a long straight feature, does not seem to be consistent with a simple 
circular motion around the cluster centre, as previously suggested in the literature. 
Two sharp bends after the straight section of the tail cast doubt on the prevailing idea in the literature that the 
long narrow tails represent trails left behind by the fast moving parent optical galaxy with respect to the cluster medium, as the optical galaxy could not have undergone 
such sharp bends in its path, under any conceivable gravitational influence of some individual galaxy or of the overall 
cluster gravitational potential. In fact the tail does not seem to have been influenced by the gravitational field of any of the 
cluster-member galaxies. The radio spectrum of the head, coinciding with the optical galaxy, is flat  ($\alpha \stackrel{<}{_{\sim}}0.4$ 
for $S \propto \nu ^{-\alpha}$),  but steadily steepens along the radio tail, with the end part of the tail showing the steepest 
spectrum ($\alpha \, {\sim}$ 4 -- 5) ever seen in any diffuse radio emission region. 
\end{abstract} 
\section{INTRODUCTION}
Radio galaxies found in clusters of galaxies often show very striking structural features not seen among those 
outside clusters. These include narrow angle tails (NATs), wide angle tails (WATs), and some other peculiar  
morphologies (see e.g., Miley 1980; O'Dea and Owen 1985; Owen and Ledlow 1997), which hopefully shed light on both the radio galaxy and its interaction 
with the cluster medium.

The interaction between the radio sources in the clusters and the cluster environment is not well understood;  
radio galaxies in the same or similar types of clusters are often found to be very different in their radio morphologies (Bliton et al. 1998; Feretti \& Giovannini 2008). 
One possible explanation for the difference in radio structures is the difference in the relative speed of the parent galaxy with respect to the surrounding 
intra-cluster medium (ICM) or differences within the ICM at the 
location of these radio galaxies. As the radio jets propagate through ICM, variations in the ICM 
density and velocity may account for the complex radio morphologies that are observed. Of course the energy budget of the 
radio galaxies in question must also be playing a crucial role while interacting with the ICM. A difference in the radio ages of the galaxies may also influence 
their appearance. 
\begin{figure*}
\includegraphics[width=\linewidth]{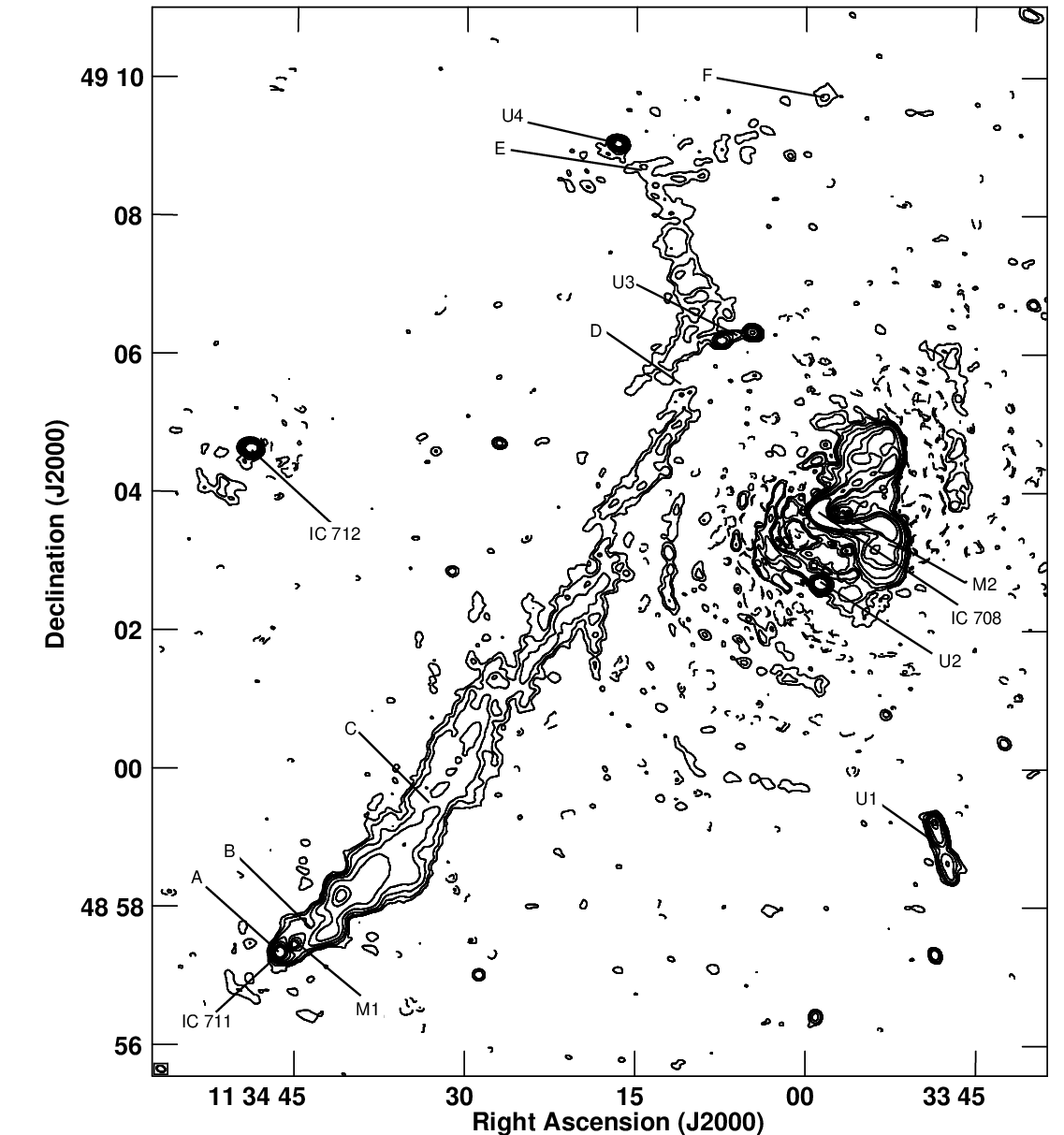}
\caption{610 MHz contour image of IC~711. The contour levels are 0.046$\times$(-4,4,8,16,32,64,128,256) mJy beam$^{-1}$. `A' is the compact head 
that coincides with the optical galaxy IC 711, whereas `B',`C',`D',`E',`F' indicate various sections of the tail. IC~708 and IC~712 are other radio galaxies in the cluster. 
`U1',`U2',`U3',`U4' are radio sources in the field which may be unrelated to the Abell cluster 1314. `M1',`M2' are actually minima (see text).}
\end{figure*}
\begin{table*}
\centering
{\bf Table 1.}\\ The observational parameters of GMRT\\
\begin{tabular}{cccccccc}
\hline\hline
\multirow{2} {*} {Frequency}   &\multirow{2} {*} {Bandwidth}  &\multirow{2} {*} {rms}&\multirow{2} {*} {Dynamic range} &\multirow{2} {*} {Primary beam}& \multicolumn{3}{c}{Restoring beam}\\ \cline{6-8} \\ 
$\nu$ (MHz)  & $\Delta\nu$ (MHz)  &   mJy/beam& achieved& $\prime$  &  $^{\prime\prime}$ & $^{\prime\prime}$ & $^\circ$ \\ 
(1)    &(2)                &(3)        &(4)    &(5)  &(6) &(7)  & (8) \\
\hline
240  &40  & 1.1  & 15    & 112 & 17.3   & 8.6       & 98      \\
610  &60  & 0.046 & 80    & 43 &7.8    & 5.2       & 71       \\
1300 &120  & 0.065& 120    &26& 3.3    & 2.7       & 43          \\
\hline
\end{tabular}
\end{table*}

In this paper, we study low-frequency radio emission from IC~711 (SDSS optical position (J2000) RA=$11^h 34^m 46^s\!\!.55$, Dec=$+48^\circ 57' 21''\!\!.9$,  r~magnitude=13.9), that displays a head-tail radio morphology. The observations were carried out with the Giant Metrewave Radio Telescope (GMRT) at 240, 610 and 1300 MHz. The radio galaxy IC~711 is a NAT, lying in Abell cluster 1314 (cluster centre (J2000) RA=$11^h 34^m 49^s$, Dec=$+49^\circ 03'\!\!.4$ (Sastry \& Rood 1971), redshift $z\sim 0.034$) that hosts another radio galaxy IC~708 in the
peculiar shape of a highly bent WAT. Abell 1314 has a number of large elliptical galaxies with the brightest of them, IC 712 (SDSS optical position (J2000) RA=$11^h 34^m 49^s\!\!.29$, Dec=$+49^\circ 04' 39''\!\!.4$,  r~magnitude=13.2), lying close to the cluster centre. 
The radio structures of the radio galaxies IC~711 
and IC~708 are very dissimilar. The radio galaxy IC~711 has a narrow long tail that may be extending beyond $\sim 800$ kpc. On the other hand IC~708 has a 
rather 'squat' radio morphology, with a wide-angle tail structure with the maximum extension an order of magnitude smaller ($\sim 80$ kpc). 
These two objects have been studied with the Westerbork telescope (Vall\'ee  \& Wilson 1976; Wilson \& Vall\'ee 1977; J\"agers 1987; 
Vall\'ee \& Strom 1988) and the Very Large Array (O'Dea and Owen 1985; Owen \& Ledlow 1997). It has been claimed (Feretti \& Giovannini 1994) that IC~712 also possesses a head-tail morphology in radio and that it is the smallest head-tail radio galaxy known. Curiously, this makes Abell 1314 unique as it then contains the longest as well as the smallest among the known head-tail radio galaxies.

Our objectives for GMRT observations
are to investigate the diffuse emission of the long tail structure of IC~711 at low frequencies, variations in its radio spectrum along the tail and to explore what further constraints can be put on the prevalent ideas in literature about the formation of the head-tail morphologies.

As will be shown, in our multi-frequency observations a break in spectrum, around 610 MHz, is seen throughout along the tail. We shall derive an expression for the minimum energy in the case of a spectral break, in order to do the minimum energy computations in the tail regions of IC~711. Further, it will be shown that the standard argument that during synchrotron radiation process the pitch angle of the radiating charge remains constant, is not conversant with the special theory of relativity. We shall provide an alternate, relativistically correct, derivation of the radiative life times of charges undergoing synchrotron losses.

\section{OBSERVATIONS}
The GMRT observations were made at 240, 610 and 1300 MHz in July 2015.
The GMRT consists of thirty 45-m antennas in an approximate `Y' shape, similar to the Very Large Array, but
with each antenna in a fixed position (Swarup et al. 1991). Twelve antennas are randomly placed within a central
1 km $\times$ 1 km square (the ``Central Square'') and the remainder form the irregularly shaped Y (6 on each arm) over a total extent 
of about 25 km. Further details about the array can be found at the GMRT website
at {\tt http://www.gmrt.ncra.tifr.res.in}. The source was observed in dual (610/240 MHz) frequency mode 
for 6 hours, and at 1300 MHz for another 6 hours, although a small percentage of data had to be edited out due to ionospheric disturbances. The observations were made in the standard fashion, with each source observation interspersed with observations of the phase calibrator. The primary flux 
density calibrator was 3C286 whose flux density was estimated to be 67.1 Jy at 240 MHz, 20.6 Jy at 610 MHz, and 15.7 Jy at 1300 MHz. The phase calibrator 
was 3C241 whose flux density was estimated to be 10.65$\pm$0.62 Jy at 240 MHz, 4.53$\pm$0.02 Jy  at 610 MHz and 2.03$\pm$0.04 Jy at 1300 MHz. 
The data analyses were done using the Astronomical Image Processing Software (AIPS) of the National Radio Astronomy Observatory.
The total flux densities of different sources or tail components of IC~711 were measured using the AIPS task TVSTAT, which allows us to pick an area of any shape. 
Error on the flux density of an extended 
component was obtained by multiplying the rms noise per beam with $\sqrt{N}$ where $N$ is the total area of the component measured in units of synthesized beams. 
\begin{figure}
\includegraphics[width=\columnwidth]{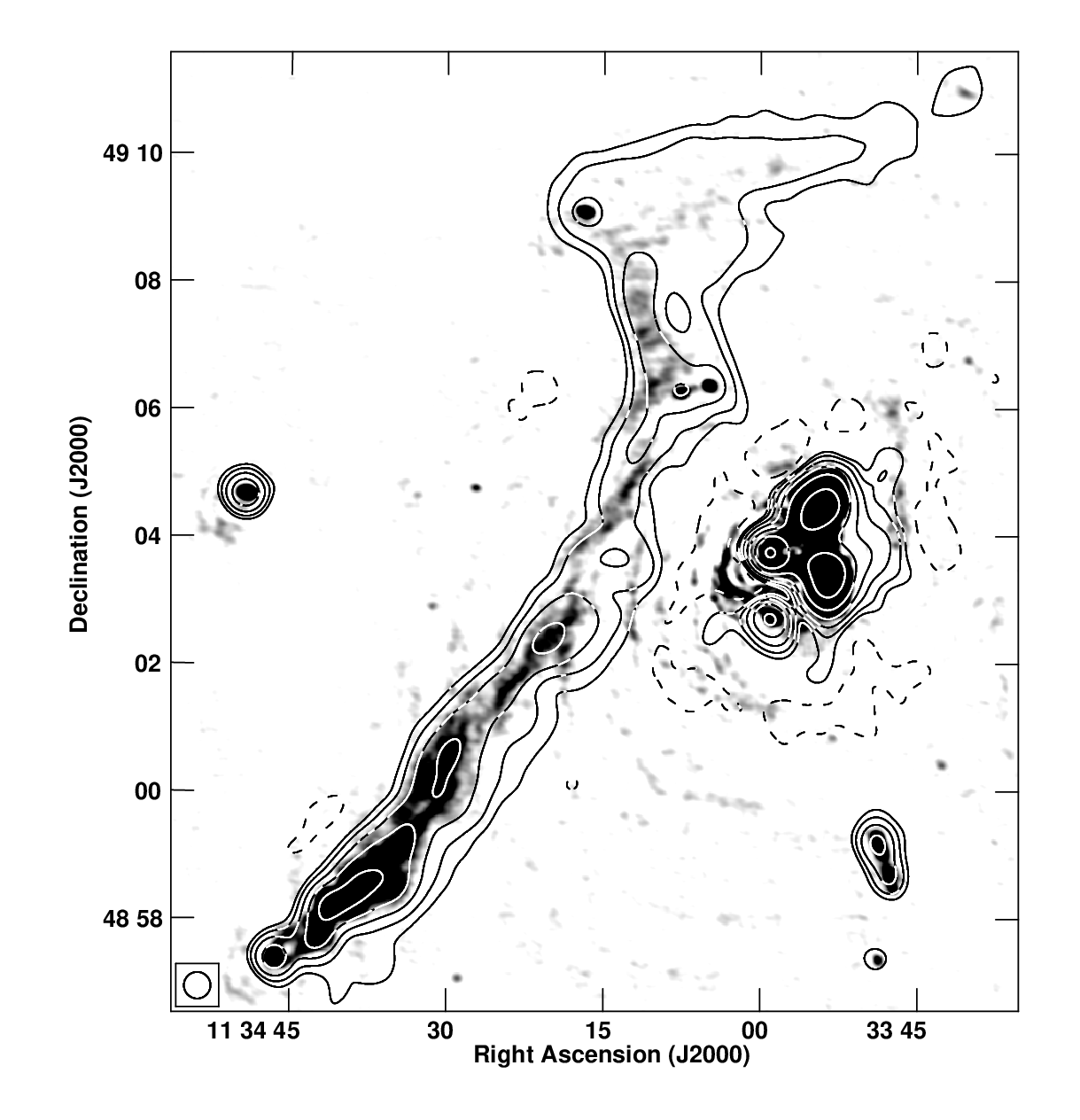}
\caption{The 150 MHz LoTSS contour map overlaid over the 610 MHz GMRT grey scale map. Grey scale flux range for the 610 MHz map is 0.1 - 0.9 mJy beam$^{-1}$.
The contour levels for the 150 MHz map are 1.46$\times$(-5,5,10,20,40,80,160,320) mJy beam$^{-1}$.}
\end{figure}

The observational parameters of the GMRT beams for different frequencies are given in Table 1, which is organized in the
following manner.
(1) Column 1-2: Frequency of observations and bandwidth in MHz. 
(2) Column 3: The rms noise in units of mJy/beam.
(3) Columns 4: Dynamic range achieved.
(4) Columns 5: The primary beam in arcmin.
(5) Columns 6-8: The major and minor axes of the restoring beam in arcsec and its PA in degrees.

\begin{figure}
\includegraphics[width=\columnwidth]{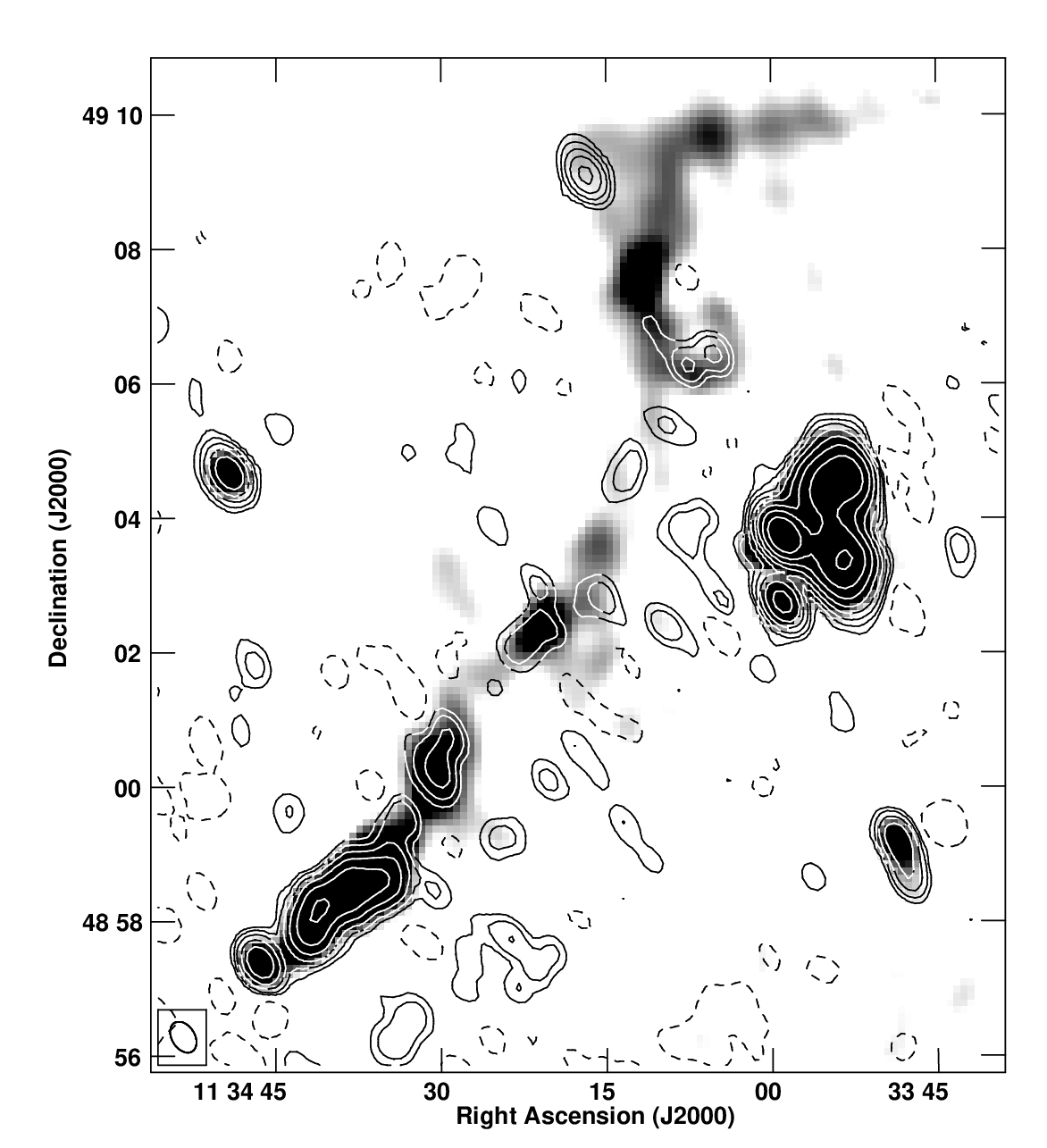}
\caption{A low resolution (30$^{\prime\prime}\times$20$^{\prime\prime}$ along  PA = $32^\circ$) 1300 MHz contour image of IC~711 superimposed on 
the grey scale TGSS ADR1 map at 150 MHz with a 25$^{\prime\prime}\times$25$^{\prime\prime}$ resolution. The grey scale flux range is 9.0 - 30.0 mJy beam$^{-1}$. The 1300 MHz contour levels are 0.23$\times$(-3,3,6,12,24,48,96,192,384) mJy beam$^{-1}$.}
\end{figure}
\section{RESULTS AND DISCUSSION}
Our detailed GMRT contour map of IC~711 at 610 MHz, with an angular resolution of $7''\!\!.8\times5''\!\!.2$, is presented in Figure~1, where  
the galaxy IC~711 shows a long radio tail up to a total angular extent $\sim 17$ arcmin, along a position angle of -71$^\circ$. This high resolution map shows that the radio tail 
along its stretch is initially straight for more than half of its length ($\sim  10$ arcmin), then there is a sharp, $\sim 120^\circ$, bend and further along, after $\sim 4$ arcmin,  
there is another similar sharp, $\sim 90^\circ$, bend pointing away from the cluster centre and the diffuse source tail continuing further for another 
$\sim  3$ arcmin. There are a number of prominent radio sources (`U1',`U2',`U3',`U4') in the field of Abell cluster 1314, but no cluster member galaxy is coincident with any of them. As we shall discuss later, these most likely are unrelated to the cluster and are merely background radio sources seen through the cluster.
 
Figure 2 shows a 150 MHz LoTSS (LOFAR Two-metre Sky Survey) contour map with a 
25$^{\prime\prime}\times$25$^{\prime\prime}$ resolution (Shimwell et al. 2017), on which we have superimposed our GMRT 610 MHz grey scale map. The two maps seem to overlap very well, but in the lower frequency LoTSS map, the tail seems to extend somewhat farther. In fact, more recent LOFAR observations of IC 711 (Wilber et al. 2019) show much more diffuse emission beyond the tail part seen in Figure~2, extending the total size of the radio galaxy to $\sim 22$ arcmin, which translates to $\sim 900$ kpc, making it the longest head-tail source seen till today.

Figure 3 shows a low resolution 1300 MHz contour map, superimposed on the TGSS ADR1 (TIFR GMRT Sky Survey -- First Alternative Data Release) 
150 MHz grey scale map with a 
25$^{\prime\prime}\times$25$^{\prime\prime}$ resolution (Intema et al. 2017). 
The original 1300 MHz contour map was convolved with a broader  beam so as to make the diffuse tail regions stand out which otherwise are not 
discernible in the high resolution map. Both the 1300 MHz and 150 MHz maps in Figure 3 have quite similar angular resolutions.
Even though the tail emission is quite weak at 1300 MHz, still it undoubtedly follows the tail emission at 150 MHz, shown as the grey scale map in Figure 3.

\begin{figure}
\includegraphics[width=\columnwidth]{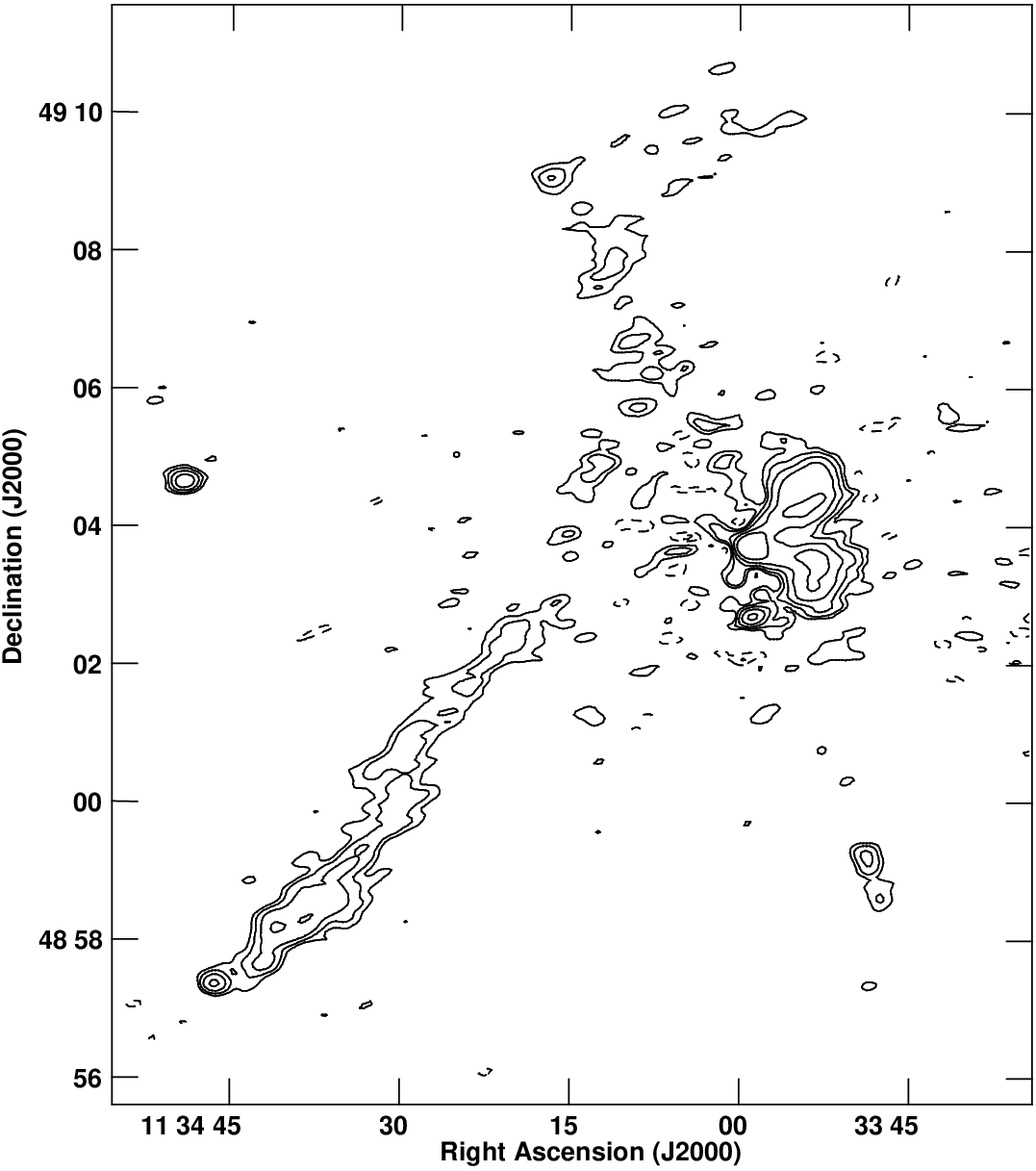}
\caption{The 240 MHz  contour image of IC~711. The contour levels are 
1.1$\times$(-3,3,6,12,32,24,48) mJy beam$^{-1}$.}
\end{figure}

Figure 4 shows the 240 MHz map of the radio galaxy IC~711, with the map showing essentially the same features as were seen in the 610 MHz map in Figure 1, though with a poorer angular resolution and a lesser sensitivity than in Figure 1. 

Figure 5 shows the optical field centred on the IC~711 galaxy (source: Sloan Digital Sky Survey), where positions of prominent optical galaxies are marked, with the 240 MHz map overlaid to show the relative radio and optical positions.
Out of the four prominent optical galaxies in the field, no radio emission has been detected from IC 709, whose projected position, otherwise, lies well within the radio emission region of IC 711. Clearer, close-up,  images of the optical galaxies are available in Vall\'ee, Bridle \& Wilson (1981). None of these cluster galaxies seems to have any direct gravitational influence on the long radio tail of  IC~711. 
\begin{figure}
\begin{center}
\includegraphics[width=\columnwidth]{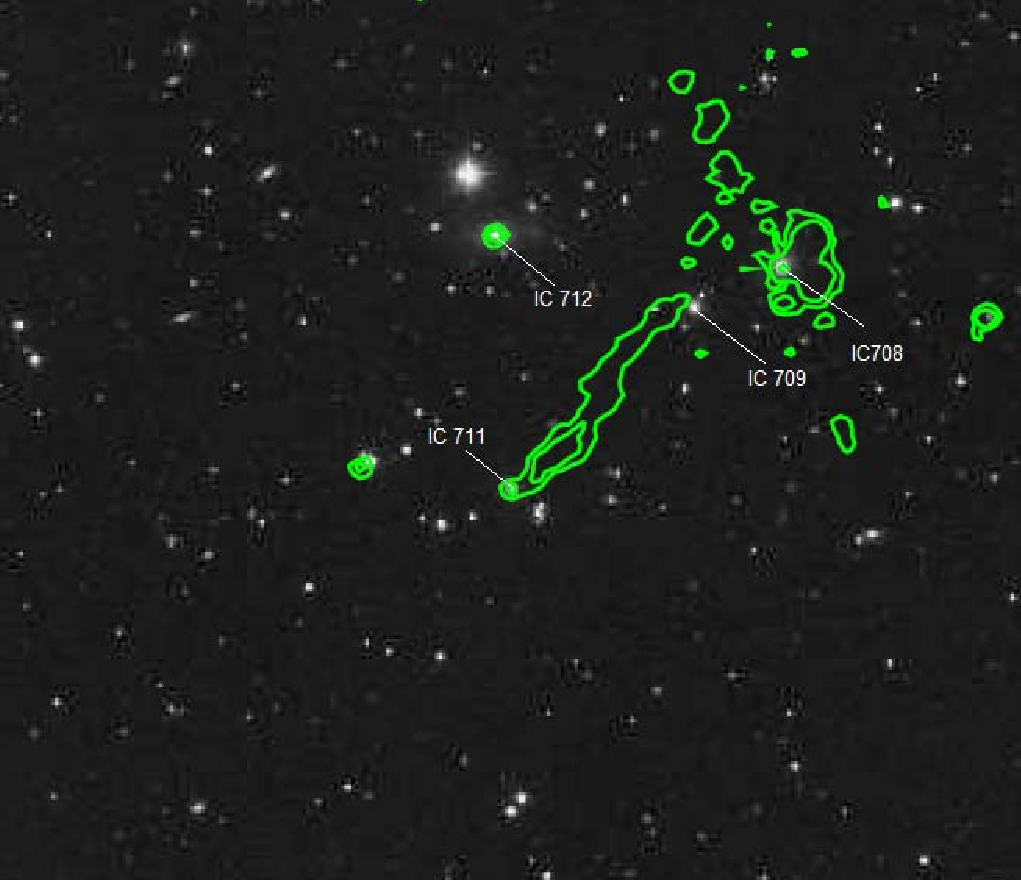}
\caption{The optical field centred on the IC~711 galaxy, with other prominent cluster galaxies  in the field, IC 708, IC 709, and IC 712 marked. The 240 MHz map is overlaid to show the relative extent of the radio emission. The contour levels for the 240 MHz map are 1.1$\times$(-3,3,6,12) mJy beam$^{-1}$.}
\end{center}
\end{figure}

The overall angular size of 
IC~711 extending to $\sim 22$ arcmin, which translates to a physical size of 
about 900 kpc (for a Hubble constant $H_{0}=70\,$km~s$^{-1}$\,Mpc$^{-1}$, the matter energy density $\Omega_m=0.3$, and 
the vacuum energy (dark energy!) density $\Omega_{\Lambda}=0.7$) at the cluster redshift of 0.034, makes it the longest head-tail radio galaxies known.
With giant radio galaxies conventionally defined as having a physical size 
$\geq 1$  Mpc for $H_{\rm o}=50$ km s$^{-1}$ Mpc$^{-1}$,  which translates to $700$ kpc or greater for $H_{\rm o}=70$ km s$^{-1}$ Mpc$^{-1}$  (Saripalli et al. 2005; Dabhade et al. 2017),
IC~711 is one of the rare examples of a non-FRII radio galaxy being the size of a giant radio galaxy.
Either of the sharp bends in the tail of IC~711 seems to occur very close to another radio source, though it does not seem  that 
the bend is physically related to that radio object. The two radio sources (`U3', `U4'), close to the bends 
(Figure 1), are most likely unrelated to the cluster as both of these radio 
sources seem to be background objects (see below), though the bends in the tail so close to two of them in sky position are very striking features.
There seems to be no direct gravitational influence of any of the prominent cluster members (i.e., optical galaxies, Figure 5) on the straight tail, particularly along its long stretch.

Figure 6 shows a higher resolution contour map of the head and the 
front part of the tail at 1300 MHz, superimposed on the grey scale map of the same region at 610 MHz. The unresolved head and twin jets 
emanating on both sides are clearly seen as modelled first by Begelman, Rees \& Blandford (1979) and Jones \& Owen (1979). We may like to point out that at 610 MHz the gap between the twin jets which is clearly seen in Figure 6 is not seen 
in Figure 1, where it instead looks like a brighter region (`M1'), but in reality is a local minima. This is because of the inability of the AIPS contour maps 
to properly distinguish between a minima (a less intense region surrounded on all sides by more brighter regions) and a maxima (a true brightness peak); these can be distinguish properly only on a grey scale/color-coded map. The head `A' coincides with the optical galaxy and has a flat radio spectrum ($\alpha \approx 0.4$; defined as S$\propto\nu^{-\alpha}$) between 240 and 1300 MHz.
\begin{figure}
\includegraphics[width=\columnwidth]{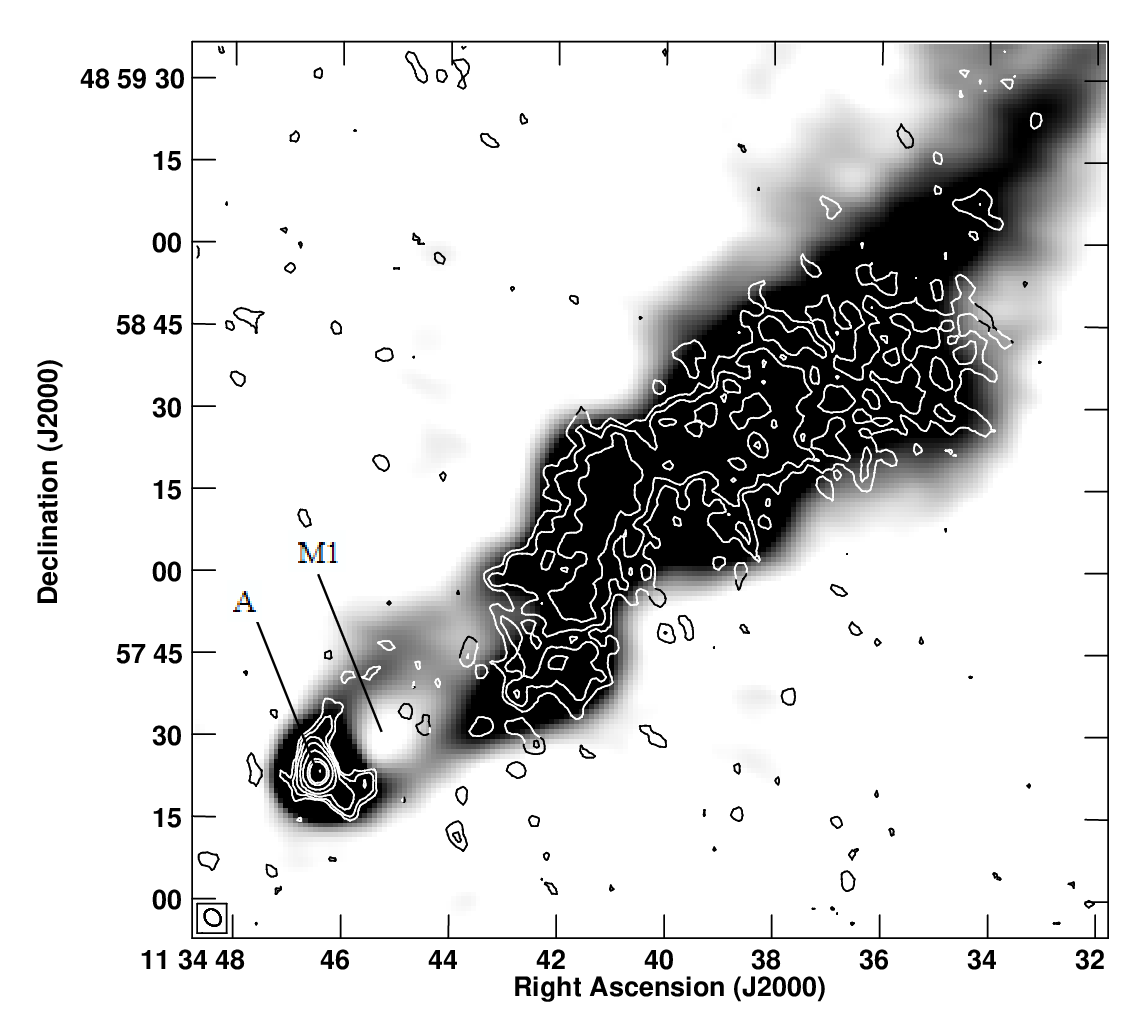}
\caption{1300 MHz high resolution map of the head and the 
front part of the tail superimposed on the 610 MHz  grey scale map of the same region. The contour levels are 0.065$\times$(-3,3,6,12,24,48,96) mJy beam$^{-1}$}
\end{figure}

\begin{figure}
\includegraphics[width=\columnwidth]{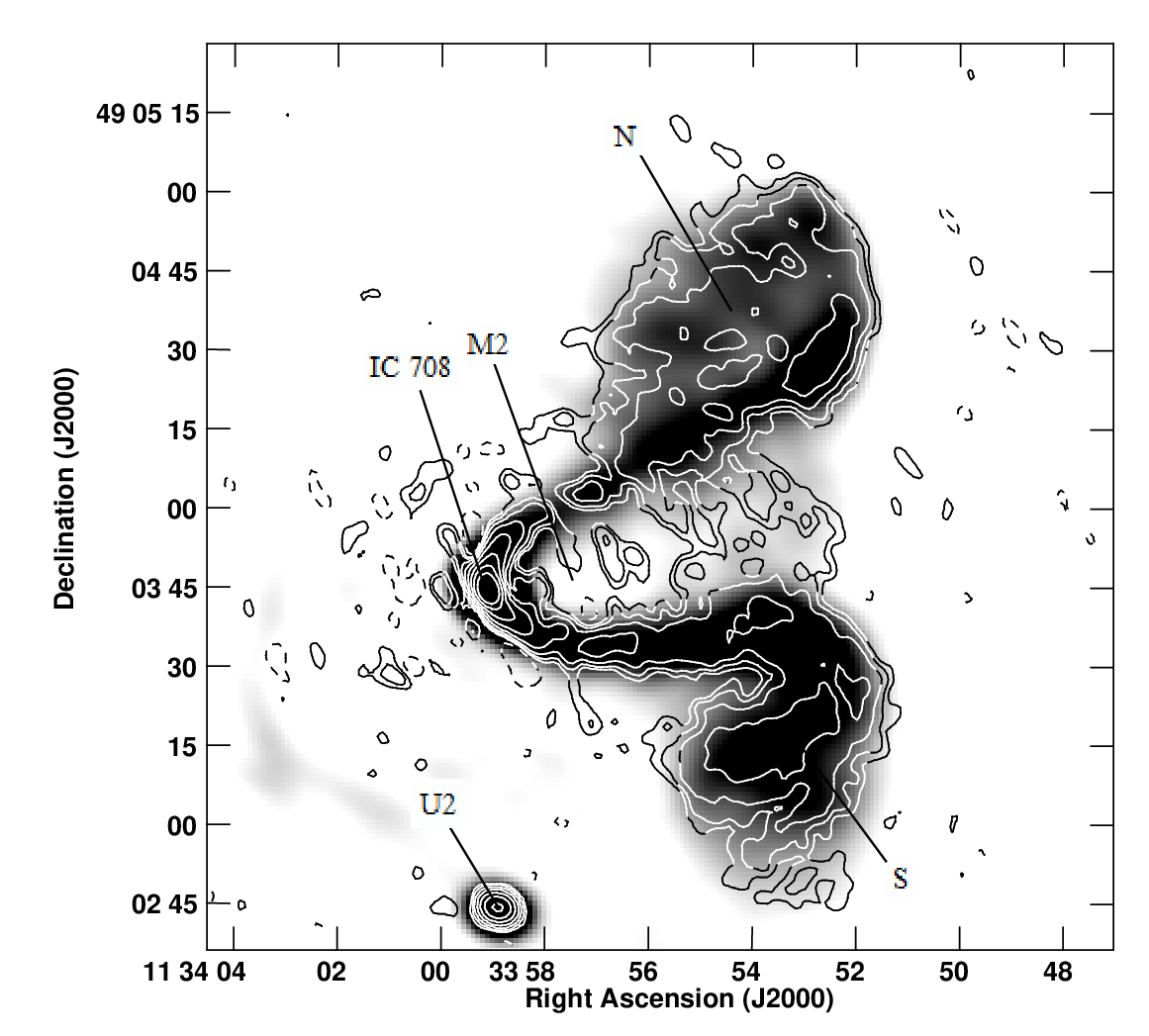}
\caption{1300 MHz high resolution map superimposed on the 610 MHz  grey scale map of IC~708. 
The grey scale flux range is 1.0 - 10.0 mJy beam$^{-1}$. The contour levels are 0.06$\times$(-4,4,8,16,32,64,128,256,512) mJy beam$^{-1}$.}
\end{figure}
\begin{table*}
\centering
{\bf Table 2.}\\ Prominent radio sources in the field of Abell 1314 unrelated to the cluster\\
\begin{tabular}{cccccccccccc}
\hline\hline
Source   & RA $\;\;\;\;\;\;$ & Dec   & 240 MHz     & 610 MHz & 1300 MHz & $\mathop{\alpha}_{240}^{610}$  & $\mathop{\alpha}_{610}^{1300}$  &  Opt. Id. & Ref.\\
        & h$\;$ m $\;$ s $\;\;\;\;\;\;$&  $\;\;\;^\circ \;\;\;\; ' \;\;\;\; ''$  & mJy         & mJy         & mJy    & &  & &   \\  
(1)    &(2)                &(3)        &(4)    &(5)&(6)&(7)&(8) &(9)&(10)    \\
\hline
`U1' &  11 33 48.1$\pm$0.2 & +48 58 49$\pm$3 &50$\pm$3 &41$\pm$2    &17$\pm$1  & 0.2$\pm$0.1 & 1.2$\pm$0.1 & 20G? & SDSS  \\
`U2' &  11 33 58.7$\pm$0.1 & +49 02 45$\pm$2  &103$\pm$5   &58$\pm$3    &34$\pm$2  & 0.6$\pm$0.1 & 0.7$\pm$0.1 & EF & -  \\
`U3'  & 11 34 06.3$\pm$0.2 & +49 06 19$\pm$2  &37$\pm$3    &18$\pm$2    &6$\pm$1   & 0.8$\pm$0.1 & 1.5$\pm$0.3 & EF & - \\
`U4'  & 11 34 16.8$\pm$0.1 & +49 09 03$\pm$2  &31$\pm$3  &25$\pm$2    &14$\pm$2  & 0.2$\pm$0.1 & 0.8$\pm$0.2 & 22G & SDSS \\ 
\hline
\end{tabular}
\end{table*}

IC708 (Figure 7) has a wide-angle tail radio structure. 
The twin jets show multiple knots and end up in two extended  
radio lobes, marked `N' (north) and `S' (south). 
This galaxy has been discussed in detail by  Vall\'ee, Wilson \&  van der Laan (1979) and Vall\'ee et al. (1981). 
Influence of the ICM seems to result in very different radio morphologies for the two neighbouring cluster members IC~708 and IC~711.
Here we want to point out that in the 610 MHz contour map (Figure 1), the apparently 
brightest feature `M2', like `M1',
is actually a minima and represents the gap between the twin jets, which is clearly seen in Figure 7. Further, the overlap between the 1300 MHz emission, 
represented by contour map, and the 610 MHz emission, shown as a grey scale map, is excellent as seen in Figure 7, both in the twin jets as well as in the extended lobes. In between the two tails there is a diffusion emission which appears to have a normal spectrum. This diffuse emission does not show a direct connection with the optical galaxy IC~708 or with the twin radio jets emanating from it, but it does appear as a fainter bridge of emission between the two extended lobes. 
\begin{figure}
\includegraphics[width=\columnwidth]{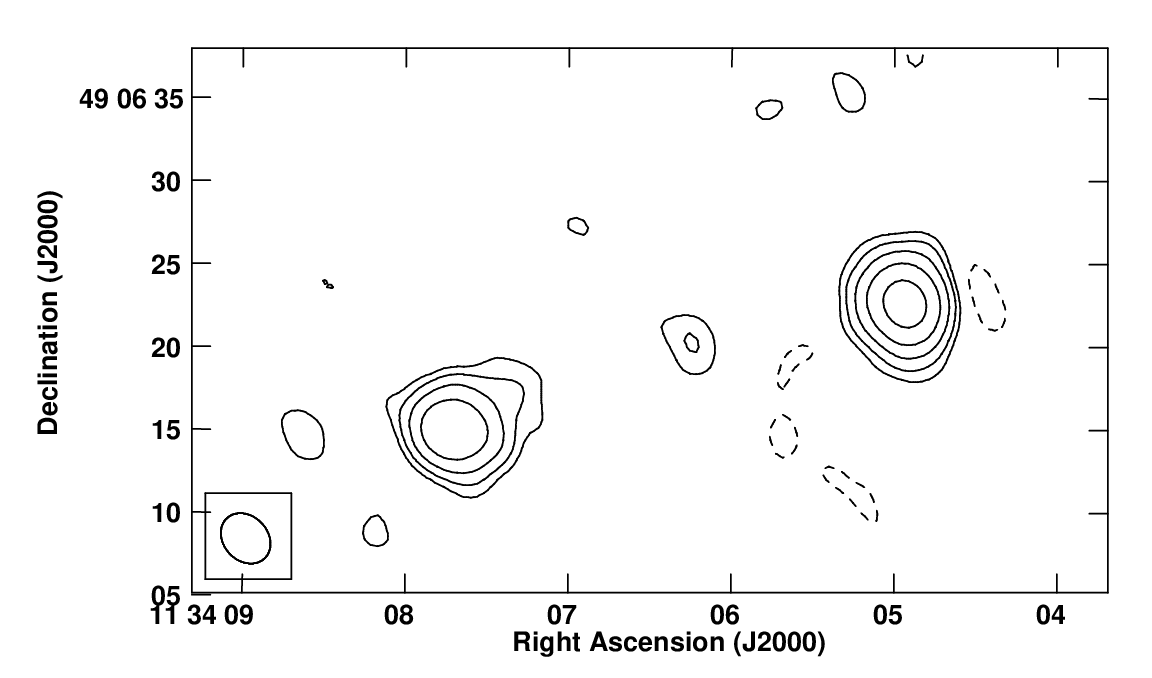}
\caption{1300 MHz high resolution map of `U3'. 
The contour levels are 0.035$\times$(-3,3,6,12,24,48,96) mJy beam$^{-1}$.}
\end{figure}

The luminosity of IC 711 at 610 MHz is $1.8 \times 10^{24}$ W Hz$^{-1}$ while that of IC 708 is $3.6 \times 10^{24}$ W Hz$^{-1}$. Thus both radio 
galaxies fall very well within the FRI luminosity range. In this case we have both a NAT as well as a WAT in the same cluster on the same side of the cluster centre, quite close to each other in sky positions.

Data for prominent radio sources in the field of Abell 1314, unrelated to the cluster, are summarized in Table 2, which is organized in the following manner. 
(1) Column 1: Radio source component. 
(2) Columns 2-3: Radio position (J2000).
(3) Columns 4-6: Flux density of the source component at the frequency  of observations in mJy.
(4) Columns 7-8: Spectral indices derived from the flux densities in columns 4-6.
(5) Column 9: Optical identification with r magnitude (G: galaxy; EF: empty field).
(6) Column 10:  Reference for optical identification (SDSS: Sloan Digital Sky Survey).

According to Vall\'ee (1988), the radio-tail hypothesis of Miley et al. (1972) can explain the shape 
of the head-tail galaxy IC~711 due to it ploughing through the dense intra-cluster gas as it moves in a simple circular 
orbit around the Abell 1314 centre where IC~712 is located (Figure 1). The low-resolution maps used by  Vall\'ee (1988) did not separate out the 
independent sources `U3' and `U4' from the diffuse tail of IC 711 (Figure 1) and that made it look like somewhat circular morphology. Vall\'ee's (1988) numerical 
simulations conform to that picture, 
where one of the main ingredient in the numerical simulation was the orbital motion of the galaxy around the cluster centre. 
But now with better angular resolutions we have not only resolved out the two sources `U3' and `U4' as not being parts of the long tail of IC~711, 
but also the long straight tail makes it unambiguously clear that it is not affected by any gravitational force of the centre galaxy IC~712. 
Moreover the extreme sharpness of tail bends after `D' and `E' also makes it clear that these bends are not result of any gravitational effect 
of any individual galaxy or even that of the overall gravitational potential of the cluster members. Nor could one think that the galaxy IC~711 
could have moved in past along a path `F'-`E'-`D'-`C' (Figure 1), as no gravitational influence on the galaxy could have made it take a path 
with such sharp bends. Sometimes a low projection angle could make small bend angles to appear much larger, but the fact that IC~711 has already the longest radio tail known, it is unlikely that it is at a low projection angle, otherwise its unprojected, physical size will be still  larger.  
\begin{figure}
\includegraphics[width=\columnwidth]{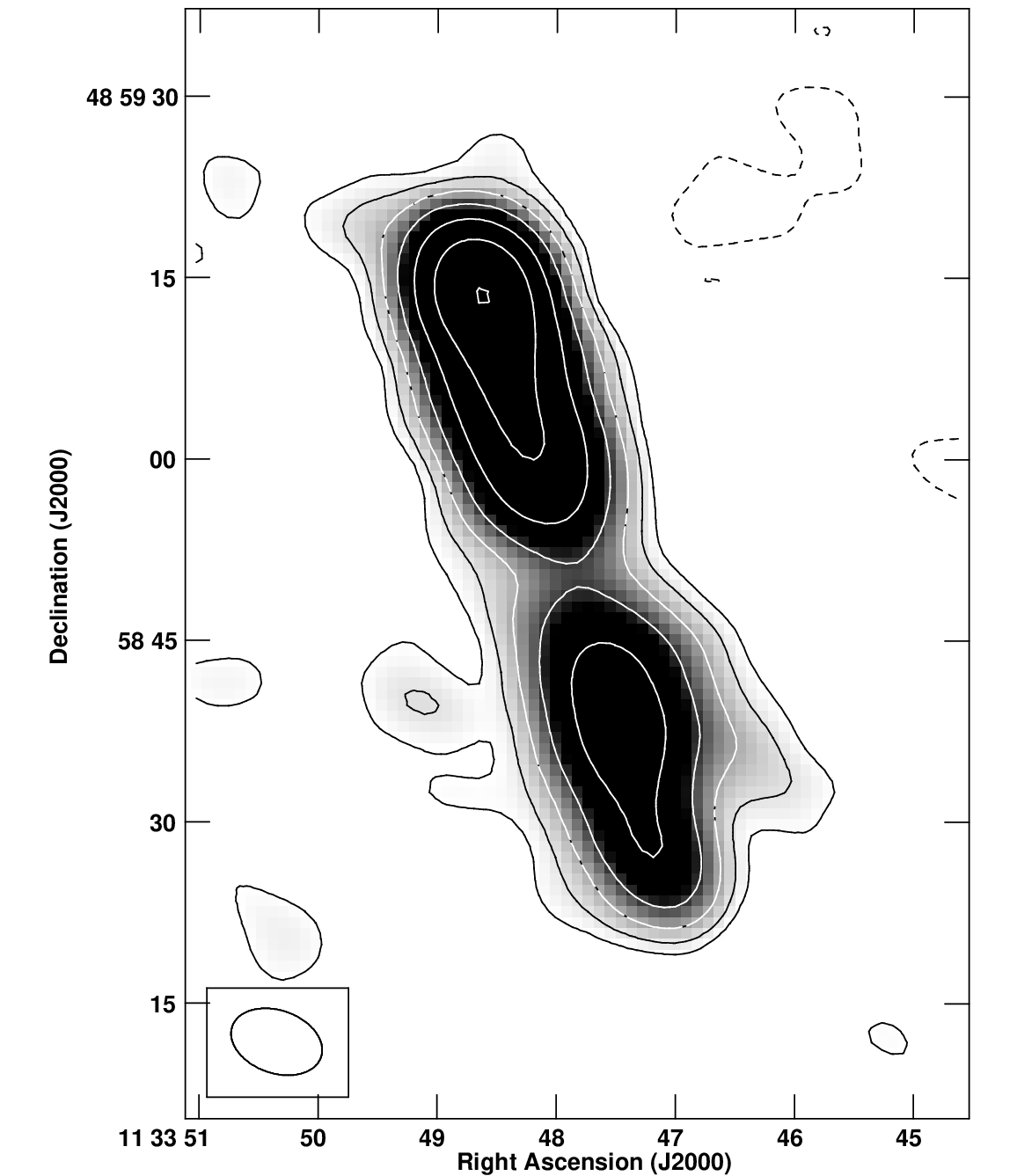}
\caption{610 MHz map of `U1'. Grey scale flux range is 0.13 - 1.3 mJy beam$^{-1}$. The contour levels are 0.0433$\times$(-3,3,6,12,24,48,96,192) mJy beam$^{-1}$.}
\end{figure}
\begin{figure}
\includegraphics[width=\columnwidth]{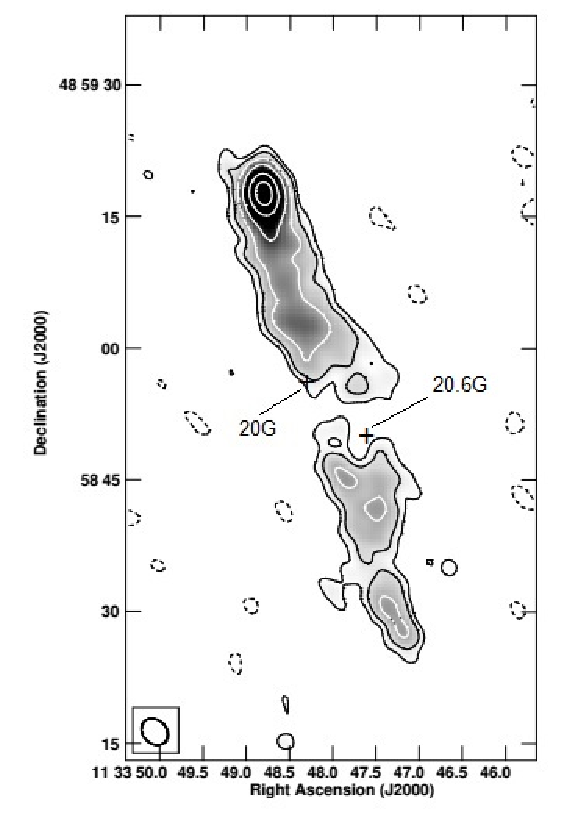}
\caption{1300 MHz high resolution map of `U1'. Grey scale flux range is 0.117 - 1.165 mJy beam$^{-1}$. The contour levels are 0.0388$\times$(-3,3,6,12,24,48,72) mJy beam$^{-1}$. Plus signs (+) mark positions of two brightest optical galaxies, with r-mag 20.0 and 20.6, lying within the radio emitting region; either of these galaxies could be the optical counterpart of the radio object `U1'.}
\end{figure}
\begin{figure}
\begin{center}
\includegraphics[width=\columnwidth]{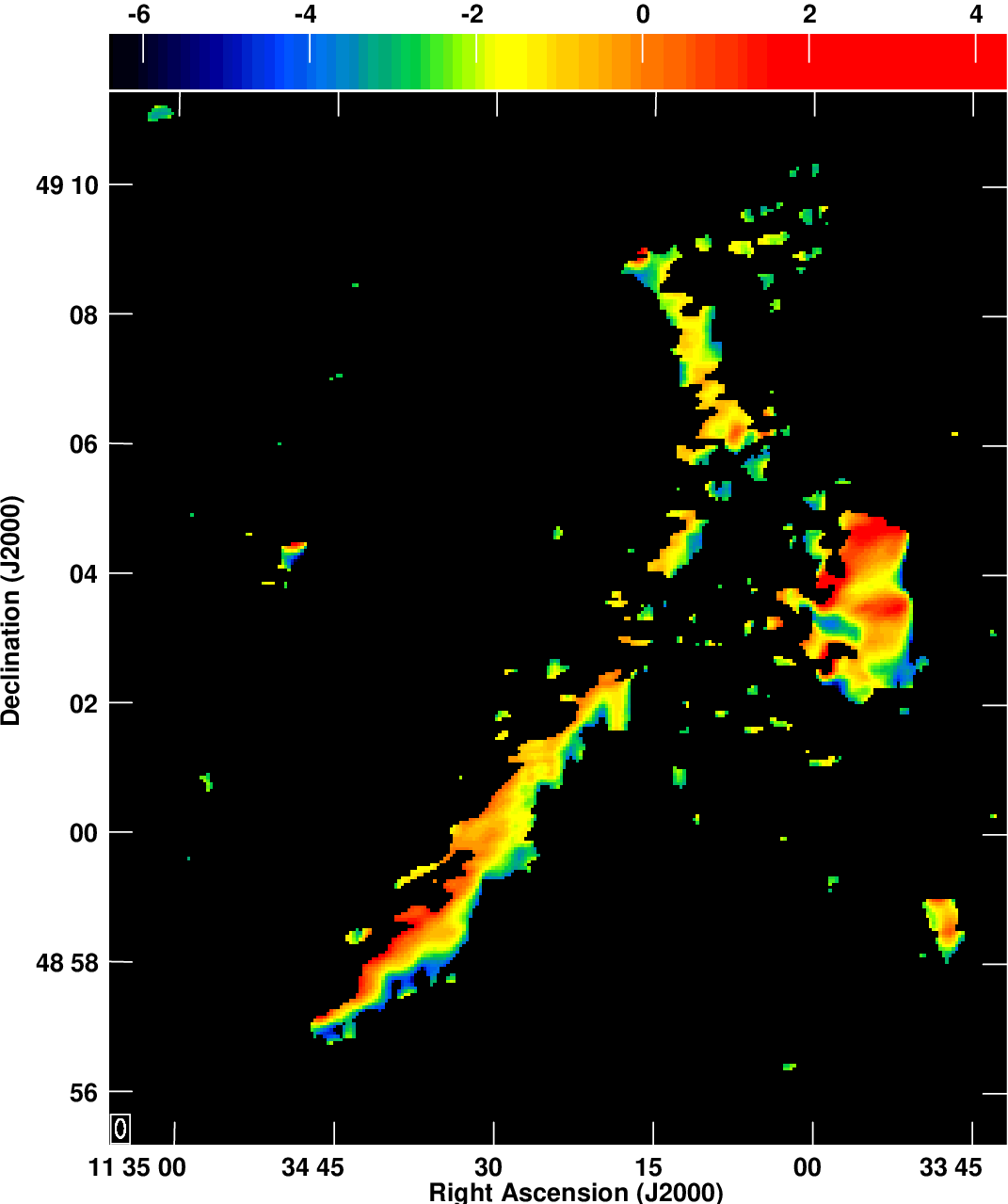}
\caption{The spectral index ($\alpha_{240}^{610}$) distribution across the radio galaxy IC~711.}
\end{center}
\end{figure}

A doubt could be raised whether the diffuse emission between `D' \& `E' or that between `E' \& `F' be unrelated to the tail of IC~711 and be some 
independent relic emission. Although such a scenario cannot be ruled out, but the continuity of the tail at `D' as well as `E', as seen especially in the 150 MHz map in Figure 2, makes such a 
possibility highly unlikely; at least we assume here that the diffuse radio emission in question is part of the tail of IC~711. 
It looks like that the sharp bends have been caused by some sort of `walls' or rather dense medium near `D' and `E' points altering the tail path (Burns 1986), though no scatter is visible where the jet 
or the tail might have encountered the `wall'. Here `U3' and `U4' could of special interest as either of them lies very close to one of the two sharp bends seen in the radio tail of IC~711 (Figure 1).  
The question whether the radio sources `U3' close to `D' as well as `U4' at `E', ostentatiously suggestive of some connection with the sharp bends, could be those `walls', seems very doubtful as from Table 2 we see that most likely these sources are not cluster members and are background sources. 
The clear double radio structure of `U3', along with possibly a core component (Figure 8), and the fact that `U4' is an unresolved flat spectrum radio source which may be optically identified with a distant galaxy (Table 2), makes both of them unrelated to IC~711 (or even to the Abell cluster 1314 itself) and their 
proximity in sky positions to tail-bends of  IC~711 seem to be mere chance coincidences. One more chance coincidence that might strike the eye is that in Figure 1 
four prominent radio sources `U1', `U2', `U3', `U4' over a separation of $\sim 12$ arcmin seem to lie, within a few degrees, in a straight line along the extended radio structure of `U1'. However, this again is a chance coincidence is corroborated by the fact that even though `U1' may appear to have extended radio structure fairly along the above mentioned straight line, the double radio structure of `U3' is almost perpendicular to that straight line.  
Another prominent radio source in the cluster field, `U2', has a normal radio spectrum (Table 2), and lies an arcmin south of IC~708 (Figure 7) but is quite clearly not associated with it. 
\begin{figure*}
\begin{center}
\includegraphics[width=\linewidth]{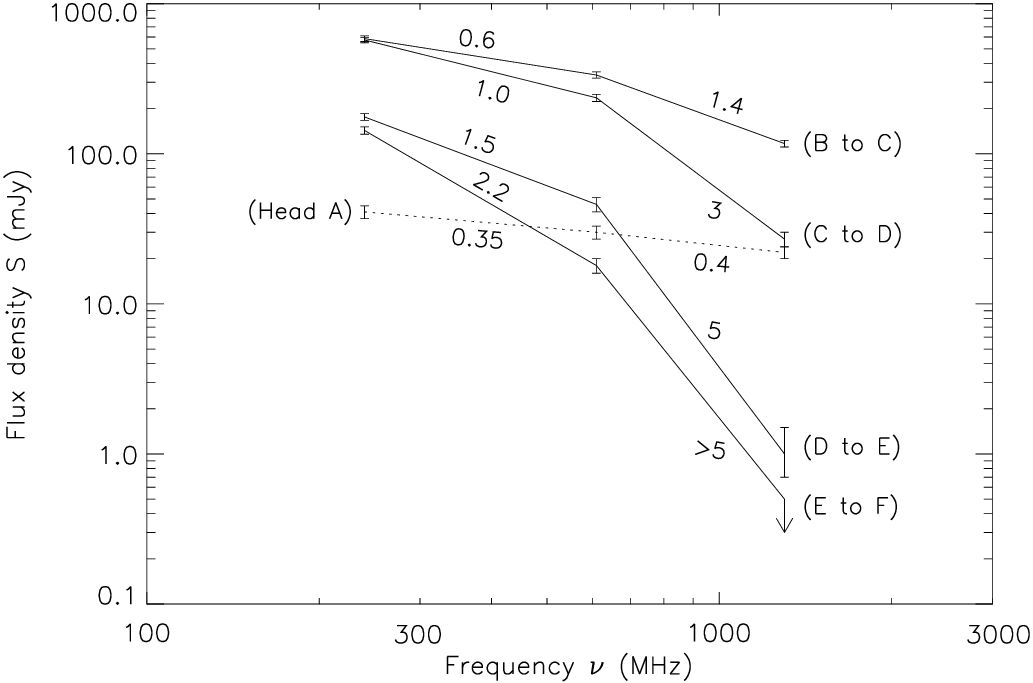}
\caption{Plots of the integrated flux densities, determined for piece-wise sections of the head-tail radio galaxy IC~711. The spectral index values between 240 and 610 MHz as well as  between 610 and 1300 MHz, calculated from straight line fits to the spectral plots of each section of the radio galaxy, are indicated above (or below) each corresponding fit of the plot. Error estimates for the spectral index values are given in Table 3. The dotted line shows a flat spectrum for the head `A', while the rest of plots show a break in the radio spectrum around 610 MHz, with the spectrum for all sections of the tail  steepening sharply above 610 MHz. The steepening in spectrum is much more conspicuous for the remote parts, 'D' to 'E' and 'E' to 'F'.}
\end{center}
\end{figure*}

`U1' is a curious source which, at 610 MHz shows the morphology of a double radio source (Figure 9). However, in spite of its resolved radio structure, the overall spectrum of radio source `U1' is quite flat ($\alpha \sim 0.2$) between 240 and 610 MHz which though steepens to $\alpha \sim 1.2$ between 610 and 1300 MHz (Table 2). When seen at a higher frequency (1300 MHz), with a better angular resolution, it looks like a narrow angle tail (NAT) 
radio source (Figure 10) with a flat spectrum head ($\alpha \sim 0.4$),  
with the diffuse tail extending to $\approx 1.0$ arcmin along a south-west direction. But then, 
in spite of the source lying in Abell cluster 1314 (redshift $\sim 0.034$, with most cluster galaxies being fairly bright with red mag $\sim 14-15$), no cluster galaxy appears to coincide with the flat spectrum head; in fact no cluster galaxy is spotted in the near vicinity of the radio emitting region. Thus it may be a curious case of a head-tail radio source 
lying in a cluster but not related to the cluster. The two brightest optical objects visible in the region are the 20 and 20.6 magnitude galaxies (Figure 10), either of which could be the optical counterpart of the radio source `U1' in case it is a double radio source, as both optical galaxies fall in between the apparent double radio components. However, for a double source, the hot spot in the northern component is of an unusually flat spectrum. In any case, `U1' then again would not be related to the Abell cluster 1314.  
It is very unlikely that `U1' is a chance coincidence of two independent radio sources, the linearly aligned radio structure of its apparent two components makes such thing a very remote possibility. It perhaps warrants a separate, thorough multi-frequency investigations, of `U1' and its relation, if any, to the Abell cluster 1314.

To examine the radio spectral index behaviour of IC 711, which is the longest head-tail radio galaxy known, 
we plotted the detailed distribution of spectral index  between 240 and 610 MHz ($\alpha_{240}^{610}$), 
along the extent of the radio emission in Figure (11). For this purpose, the 610 MHz map was reduced to the 
same angular resolution as that of the 240 MHz map, after applying the primary beam corrections. 
As one would expect, the spectra steepens in the remote parts (`E' to `F') of the tail, with $\alpha_{240}^{610} \rightarrow 2$. 
But what may be more curious is that there appears to be a gradual steepening of the spectrum across the width of the tail as 
we move from upper (northern) edge to the lower (southern) edge of the tail. Though a shift in the relative position of the two 
images could cause such an artefact to appear, it is not clear what could have caused such a shift in relative positions. 
Still, we cannot be sure how much of this gradual steepening could be genuine as a somewhat similar trend in the spectral map 
is visible for other cluster galaxies (see, e.g., IC 708).

For a more quantitative study of the spectral variation along the tail of this longest head-tail galaxy known, apart from the unresolved head at `A', we divided the radio emission into various sections between points `B', `C'. `D', `E' and `F' (Figure 1) and determined the integrated flux density for piece-wise sections at all three frequency bands and therefrom determined the spectrum index values between 240 and 610 MHz ($\alpha_{240}^{610}$) as well as between 610 and 1300 MHz ($\alpha_{610}^{1300}$). The flux-density values, along with their estimated errors, for the head at `A' and the other piece-wise sections, are plotted in Figure 12. 
\begin{table*}
\centering
{\bf Table 3.}\\ Spectral index, minimum energy, equipartition magnetic field and radiative life-time estimates\\
\begin{tabular}{cccccccccccc}
\hline\hline
Components  & 240 MHz     & 610 MHz & 1300 MHz & $\mathop{\alpha}_{240}^{610}$  & $\mathop{\alpha}_{610}^{1300}$  &  $\psi_x$ & $\psi_y$ & $U_{\rm min}$ & 
$B_{\rm eq}$  & $\tau_{rad}$ \\
            & mJy         & mJy         & mJy    & &  & $''$ & $''$ & $10^{-13}$ erg cm$^{-3}$ & $\mu$G &  $10^{8}$  yr  \\  
(1)    &(2)                &(3)        &(4)    &(5)&(6)&(7)&(8)&(9)&(10)&(11)      \\
\hline
IC 711 `A'       &41$\pm$4   &30$\pm$3    &22$\pm$2   & 0.35$\pm$0.15  & 0.4$\pm$0.2 & - & - & - & - & - \\
IC 711 `B' to `C'   &585 $\pm$ 25  &335 $\pm$ 16 &117 $\pm$ 6  &0.6$\pm$0.1 &1.4$\pm$0.1  & 167 & 111 & 1.3 & 1.2& 1.5 \\
IC 711 `C' to `D'   &573 $\pm$ 23 &236 $\pm$ 13  &27 $\pm$ 3 &1.0$\pm$0.1 &3$\pm$0.2 & 445   &   56 &  1.4 & 1.2& 1.5 \\
IC 711 `D' to `E'   &176 $\pm$ 9  &46 $\pm$ 5    &  $1^{+0.5}_{-0.3}$   &1.5$\pm$0.1 &5$\pm$0.7 & 222  &  28 & 2.4 & 1.6& 2.5 \\
IC 711 `E' to `F'   &143 $\pm$ 8  &18 $\pm$ 2 & $\stackrel{<}{_{\sim}}$ 0.5 &2.2$\pm$0.2 & $\stackrel{>}{_{\sim}}$ 5 & 167 & 28 & 2.6 &1.7& 2.5 \\
IC 708 `N'  &636 $\pm$ 30  &372 $\pm$ 20 & 239$\pm$ 15 &0.6$\pm$0.1 & 0.6$\pm$0.1 & 68 & 47 & 7.7 &2.9& 0.5 \\
IC 708 `S' &577 $\pm$ 28  &383 $\pm$ 18 &  228$\pm$ 14 &0.4$\pm$0.1 & 0.7$\pm$0.1 & 66 & 47 & 7.4 &2.8& 0.5 \\
IC 708        &2072$\pm$60  &1363$\pm$42  &791$\pm$24 & 0.45$\pm$0.05 & 0.7$\pm$0.1 & - & - & - & - & - \\
IC 712        &69$\pm$4    &46$\pm$3    &24$\pm$2  & 0.4$\pm$0.1 & 0.9$\pm$0.1 & - & - & - & - & - \\
\hline
\end{tabular}
\end{table*}

The flux-density and spectrum index values, along with their estimated errors, can be read from Table 3, which is organized in the following manner. 
(1) Column 1: Radio source component. 
(2) Columns 2-4: Flux density of the source component at the frequency  of observations in mJy.
(3) Columns 5-6: Spectral indices derived from the flux densities in columns 2-4.
(4) Columns 7-8: Angular size of the component along two perpendicular axes.
(5) Column 9: Minimum energy value.
(6) Column 10: Equipartition magnetic field value.
(7) Column 11: Radiative life time of the source component.

The spectral index of the head (`A') of IC 711 is flat between 240 and 1300 MHz (Table 3), in agreement with the flux densities measured from the VLA observations, $25.92\pm 0.03$ at 1.5 GHz, $24.30 \pm 0.02$ at 4.5 GHz, and $23.23 \pm 0.01$ at 7.5 GHz, showing a flat spectrum ($\alpha = 0.07$, Baldi, Capetti \& Giovannini 2019), where the VLA higher resolution maps show  the twin jets to be  highly asymmetric in the innermost regions, especially at 7.5 GHz. Our somewhat higher flux density values at 240 and 610 MHz (Table 3) could be because at the lower resolutions some part of the twin jet emission might have got included in our measurements, though we still find the spectral index to be flat with $\alpha \stackrel{<}{_{\sim}} 0.4$. 

However, the spectral index of the tail steadily steepens as we move away from the head (Table 3, Figure 12). The closer parts of the tail (`B' to `C' and `C' to `D') are having still 
a normal spectra ($0.6\stackrel{<}{_{\sim}}\alpha \stackrel{<}{_{\sim}}1$) between 240 and 610 MHz, but the spectra of the remote parts of the tail (`D' to `E' and `E' to `F') become quite steep ($ 1.5\stackrel{<}{_{\sim}} \alpha \stackrel{<}{_{\sim}}2.2$) at these frequencies. That the spectrum between 240 and 610 MHz steepens along the tail from ($ {\sim} 0.7$ to $ {\sim} 2.1$)  is also seen by Sebastian, Lal \& Rao (2017). 
However,  between 610 and 1300 MHz, the spectra of all parts of the tail is much steeper, 
thereby indicating a break in the spectrum at or around 610 MHz. In fact, at frequencies higher than 610 MHz, these end parts of the tail show the steepest spectrum ($\alpha \stackrel{>}{_{\sim}}5$), ever seen in any diffuse radio emission region, steeper than even the extremely steep ($\alpha \stackrel{>}{_{\sim}}2$) radio halos or relic radio emission regions  hitherto known (see, e.g., Slee et al. 2001; Pizzo \& Bruyn 2009; Cohen \& Clarke 2011; Bonafede et al. 2012). The closer parts of the tail ('B' to 'C' and 'C' to 'D') may have  
break frequency not very far from 610 MHz, but the remote parts of the tail ('D' to 'E' and 'E'  to 'F') may have a break frequency more closer to 240 MHz. But since 
we have the detailed flux density measurements of the different parts of the tail at only these three frequencies, we take 610 MHz as the break frequency 
for our estimates of the minimum energy as we find the spectra differ above and below this frequency (Figure 12). 

We may mention here that we have selected tail sections which are 
many times larger in angular size than the broadest beam we have used, then the flux density 
estimates are not very much dependent upon the beam size and shape. Then it is not 
very essential to convolve the images to a common resolution.
We explicitly checked this procedure by determining flux density for various tail sections 
both for high and low resolutions (1300 MHz) maps and got quite the same answers 
(not that we expected it to happen otherwise). Of course all images are primary beam corrected. 
Therefore we are quite confident that our flux-density and spectral values are reliable.
Further, the low flux-density beyond `D' or `E' at 1300 MHz, resulting in  $\alpha\stackrel{>}{_{\sim}} 5$,
is not due to any missing small spatial frequencies, otherwise this would have resulted in similar 
low flux density and thus rather steep $\alpha$ between `B' and `C' or between `C' and `D' as well. 
Moreover, maps by  Wilson \& Vall\'ee (1977) and by J\"agers (1987) at 0.6 GHz and 1.4 GHz  
show the gross behaviour of the radio tail to be quite similar, though with somewhat lesser details, 
to our maps here at 610 MHz and 1300 MHz; there is hardly any diffuse emission visible at 1.4 GHz beyond point `D'. 
We also estimated the flux densities in the tail parts 'D' to 'E' and 'E to 'F' in the 1.4 GHz NVSS map (Condon et al. 1998), 
that turned out to be $1.4\pm 0.4$ mJy and $1\pm 0.5$ mJy respectively, 
which are quite consistent with our values  of the flux-densities (Table 3) and the thereby estimated very high spectral 
index ($\alpha \, {\sim}$ 4 -- 5) in these tail parts.
\subsection{MINIMUM ENERGY ESTIMATES}
An expression for minimum energy density of a synchrotron radio source component for a single power law index is  available in literature (Pacholczyk 1970; Miley 1980; Singal et al. 2004). In Appendix A we have derived the minimum energy density, $U_{\rm min}$, for a synchrotron radio source component having a {\em spectral break}  
\begin{eqnarray}
\nonumber 
U_{\rm min}& =  4.5 \times 10^{-10} \left[ \frac {(1+z)^{3.5} \: S_{0}\:  \nu_{0} ^{0.5}\: }  {\psi_{x} \psi_{y}\: s\: } 
\left\{\frac {(\nu_1/\nu_0)^{-\alpha_1+0.5} -1 }{\alpha_1-0.5}\right.\right.\\
& \left.\left. 
-\frac {(\nu_2/\nu_0)^{-\alpha_2+0.5} -1}{\alpha_2-0.5}  \right\}\right] ^{4/7} 
{\rm erg\: cm}^{-3} \;.
\end{eqnarray}
Here $z$ is the redshift, $S_{0}$ (Jy) is the flux density at the break frequency 
$\nu_{0}$ (GHz), $\alpha_1$ and $\alpha_2$ are the spectral indices before and after the break, with $\nu_{1}$ and $\nu_{2}$ (GHz) as the lower and upper cut off
frequencies presumably for the observed radio spectrum, $\psi_{x}$ and $\psi_{y}$ (arcsec) represent the size of the source component along its major and minor 
radio axes, 
and $s$ is the path length through the component along the line of sight in kpc. 
In the case of  IC~711, with its long head-tail radio morphology, we take it to 
be a bent cylinder and take its depth to be equal to its minor radio axis and 
thus $s$ is calculated from $\psi_{y}$, the width of the tail.
At a redshift of 0.034, for $H_{\rm o}=70$ km s$^{-1}$ Mpc$^{-1}$ it scales to 0.68 kpc per arcsec.
We have assumed an equal energy 
distribution between the electrons and the heavy particles ($k=1$), taken 
the volume filling factor to be unity ($\phi=1$) and also assumed the
pitch angle to be $90^\circ$. Apart from other things these are consistent with our objective of finding at least a minimum value estimate of the energy density 
in the radio source component, in the absence of any other specific information.

The equipartition magnetic field $B_{\rm eq}$ is calculated from Equation~(\ref{eq:64q1})
\begin{eqnarray}
\label{eq:64q}
B_{\rm eq} \approx B{\rm min} =3.3 \: U_{\rm min}^{0.5}\; G.
\end{eqnarray}
The estimates of the minimum energy densities and equipartition magnetic fields for different parts of the tail are given in Table 3.

Here we have integrated the radio emission from 100 MHz to 10 GHz. It should be noted that the upper cut-off frequency value is less critical for the minimum energy estimates. From that it might appear that to have separate estimates below and above the spectral break might not make any difference. To check that we made minimum energy estimates in the standard approach without using the break and we find estimates could differ on the average as much as 15 percent. We have avoided using Beck \& Krause(2005) formulation for minimum energy estimates due to two reasons. Firstly a power law between a minimum and a maximum frequency is an observational result (at least in principle) and then one may keep the energy interval floating depending upon the minimization process. But to assume an energy interval to begin with is more artificial. Moreover the procedure adopted by Beck \& Krause (2005) uses a lower energy limit which is below even the rest mass energy of the charge (their Eq. (A1)), which certainly is not a correct thing to do. 

Surprisingly, minimum energy estimates appear somewhat larger in remote parts of the tail, i.e., tail parts lying between `D' and `E' or between `E' and `F' (Figure 1). This is because the spectrum in each of these cases is already quite steep between 240 and 610 MHz and most likely the break frequency is much lower than our assumed 
610 MHz. But we do not have any measurements of the flux densities of these tail parts at still lower frequencies. The rather steep spectral index values between 240 and 610 MHz results in unusually larger number of electrons at lower cut-off frequency limit and thus much larger $U_{\rm min}$ estimates. On the other hand for IC~708 `N' and `S' components or even of the whole radio source the spectrum is not steep (Table 3). In fact from the 5 GHz map of Vall\'ee \& Wilson (1976) it seems the break frequency for 
IC~708 `N' and `S' components is $\stackrel{>}{_{\sim}}5$ GHz.  Our integrated 1300 MHz flux density measurements 791$\pm$24 mJy of IC~708 are consistent with the measurements 835 $\pm$80 mJy at 1.4 GHz by Wilson \& Vall\'ee (1977). Also our 610 MHz flux density value (1363$\pm$42 mJy) is equally consistent 
with the value 1432$\pm$70 mJy at 610 MHz, as measured by Vall\'ee \& Wilson (1976).

Radio tail of IC~711 not only seems to be straight most of its length (except for the two sharp bends mentioned earlier), it also seems to remain confined to quite a 
narrow width without any lateral free-expansion of the tail material, perhaps confined by the  pressure of the ICM. The minimum energy of the tail  may be used  
to estimate the pressure in the ICM, whose emissivity is otherwise not directly detectable, 
by assuming that the diffuse parts of the radio tail are in pressure equilibrium with it. The pressure of such a hot, diffuse, non-relativistic ICM could be 
thus estimated. 
Our estimate for the pressure in the tail part yields $P \sim 2 \times 10^{-13}$ dyne cm$^{-2}$ under the assumption of minimum energy in the radio tail and its 
pressure balance with the ICM in the cluster. 
\subsection{RADIATIVE LIFE TIMES} 
In synchrotron radiation there is a paradox whether or not the pitch angle of a radiating charge varies (Singal 2018). 
The conventional wisdom is that the pitch angle does not change during the radiation process. 
The argument put forward in the literature is that the radiation is confined to a narrow cone of angle $1/\gamma$ 
around the instantaneous direction of motion of the charge, (Jackson 1975; Pacholczyk 1970; Rybicky \& Lightman 1979), therefore from conservation of momentum,  
the radiation reaction on the charge would be in a direction opposite to its instantaneous velocity vector.
As a consequence the direction of motion of the charge should not change, implying a constant pitch angle of the charge motion (Kardashev 1962;  Pacholczyk 1970). The accordingly derived formulas for 
energy losses of synchrotron electrons in radio galaxies are in use for the last 50 years. However, using the velocity transformation from the special theory of relativity, we show in Appendix B that the pitch angle of a radiating charge varies. The velocity 
component parallel to the magnetic field does not get affected, on the other hand magnitude of the perpendicular component reduces due to radiative losses, therefore a change in the pitch angle takes place.

The standard formula used in literature for the synchrotron power loss 
(Kardashev 1962; Pacholczyk 1970; Rybicki \& Lightman 1989) is
\begin{equation}
\label{eq:30.1}
\frac{{\rm d}{\gamma}}{{\rm d}t} = -{\eta}\sin^2 \theta\;\gamma^2.
\end{equation}
Here ${\eta}=2e^4 B^2/(3m_{\rm o}^3c^5)=1.94 \times 10^{-9}B^2\; {\rm s}^{-1}$.
Something is amiss in this power loss formula can be readily seen for $\gamma=1$ case, where all that the charge has got is only its rest mass energy but the formula still yields a finite power loss rate.

If $\gamma_{\rm 0}$ represents the initial energy  at $t=0$, then under the supposition that the pitch angle is a constant of motion, from Equation~(\ref{eq:30.1}) the energy of the radiating charge at $t=\tau$ becomes 
\begin{equation}
\label{eq:30.4}
\gamma= \frac{\gamma_{\rm 0}}{1+\eta\sin^2 \theta\;  \tau\; \gamma_{\rm 0}}\;.
\end{equation}
It follows from Equation~(\ref{eq:30.4}) that energy of the electron reduces to half its value in  time $\tau_{1/2} = 1/(\eta\sin^2 \theta \;\gamma_{\rm o})$ 
(Kardashev 1962; Kellermann 1964; van der Laan \& Perola 1969; Pacholczyk 1970; Miley 1980).  

From Equation~(\ref{eq:30.4}), one can readily see that while it might work   
for a highly relativistic charge ($\gamma\gg1$), it cannot be true in general. 
This is because from Equation~(\ref{eq:30.4}) $\gamma\rightarrow 0$ when $\tau\rightarrow\infty$, while we know that $\gamma\ge1$ always. Actually an 
approximation $\beta \approx 1$ has been used in Equation~(\ref{eq:30.1}); 
the exact equation for the energy loss rate (Melrose 1971; Jackson 1975; Longair 2011) is
\begin{equation}
\label{eq:30.2}
\frac{{\rm d}{\gamma}}{{\rm d}t} = -{\eta}\sin^2 \theta\;\beta ^2\gamma^2.
\end{equation}

Using the correct expression for the radiative power loss rate, and also taking the pitch angle changes into account, we derive 
in Appendix B the exact expression for energy losses for a charge of any given $\gamma$ and accordingly derive the synchrotron life-times of a radiating charge.

Assuming that the spectral break lies around $\nu^*$, the synchrotron life-times of radiating 
electrons are calculated from Equation~(\ref{eq:30.2i}) as  
\begin{equation}
\label{eq:64r}
\tau_{syn} \approx 1.06 \left(\frac{B_\perp}{\mu{\rm G}}\right)^{-1.5} \left(\frac{\nu^*}{\rm GHz}\right)^{-0.5}\;\;{\rm Gyr}\;.
\end{equation}
Since we are using the equipartition magnetic field determined from Equation~(\ref{eq:64q}), where pitch angle was assumed to be $90^\circ$, 
to be consistent, we put $B=B_\perp$ here also. Synchrotron life-times turn out to be about $0.7-1$ Gyr, taking $\nu^*=0.61$ GHz. 

However, following van der Laan \& Perola (1969) (also see Slee et al. 2001) if we include the inverse-Compton losses from the Cosmic Microwave Background Radiation (CMBR) 
with energy density $U_R$ or the equivalent magnetic field, $B_{R}=\sqrt {8\pi U_R}$ we can write the radiative life-times as 
\begin{eqnarray}
\nonumber
\tau_{rad}& \approx &1.06 \left(\frac{B_\perp}{\mu{\rm G}}\right)^{0.5} \left[\left(\frac{B_\perp}{\mu{\rm G}}\right)^2
+\frac{2}{3}\left(\frac{B_R}{\mu{\rm G}}\right)^2\right]^{-1} \\
\label{eq:64s}
&&\times\left(\frac{\nu^*(1+z)}{\rm GHz}\right)^{-0.5}\;\;{\rm Gyr}.
\end{eqnarray}

With $U_R=4.2\times 10^{-13} (1+z)^4 = 4.8\times 10^{-13}$ erg cm$^{-3}$ and the equivalent magnetic field, $B_{R}= 3.25 (1+z)^2 = 3.5 \:\mu$G, the 
radiative life-times are about an order of magnitude lower $\sim 1.5\times  10^{8}$ yr (Table 3). For estimating the radiative life times of the tail parts between `D' and `E' or between `E' and `F', we have taken the break frequency to be 240 MHz, which we think is a more realistic value than 610 MHz. We should also add here that since the inverse Compton losses affect all electrons independently of their pitch angle with respect to the local magnetic field, even if the synchrotron losses tend to produce an anisotropic pitch angle distribution as time proceeds,
in weak magnetic field cases the inverse Compton losses may arrest this tendency to some extent and thereby tend to keep the electron population isotropic.

For the galaxy IC 711 moving within the cluster with a 
velocity $\sim 1200$ km s$^{-1}$ (Coleman et al. 1976), it would have taken $\stackrel{>}{_{\sim}} 7.5\times 10^8$ years (dynamical age) to cover the total length of the radio emitting regions. The overall source must be in the sky plane, otherwise the actual physical length of the tail would be even longer. 
Coleman et al. (1976) have estimated that IC 711 radio tail may be making an angle $\eta \sim 20^\circ$ with respect  to the sky plane and then its physical 
length will be about $1/\cos \eta \sim 10\%$ longer than that the projected length observed in the sky plane, making the dynamical age $\stackrel{>}{_{\sim}} 8\times 10^8$ years. Sebastian et al. (2017) have calculated the dynamical age of the IC 711 radio galaxy to be $\stackrel{>}{_{\sim}} 10^9$ years, an order of magnitude larger than the radiative life-times of electrons.
Now, according to the prevailing idea in the literature, the 
tails in heat-tail radio galaxies represent trails left behind by the parent optical galaxy in the cluster medium, then the electrons in these remote parts of the tail are still radiating after ${\sim} 10^9$ years, much longer than their radiative life times $\sim 10^{8}$ years, implying some re-acceleration occurrence in these tail parts. Here it should be noted that at about a billion years further back, the CMBR  
energy density and the equivalent magnetic field would be stronger which would make the inverse-Compton losses from the CMBR to be still higher and consequently the radiative life-times to be shorter about 10-20 percent than those in Table 3.

In our calculations, we have not considered injection of any new particles as it is unlikely that energy could 
be continually supplied by the head (main galaxy) to its much lagging behind tail parts, especially through the two sharp bends. Recently Wilber et al. (2019), from the 
0.3-2.0 keV X-ray band image of Abell 1314 obtained with XMM-Newton, inferred that 
the middle portion of the radio tail, where the X-ray brightness is highest along the tail, may indicate possible local re-acceleration or compression. However, it may not be true everywhere, especially in the remote parts of the tail, where our main interest lies because of the  observed extremely steep spectrum.

As explained earlier, for calculating radiative life times for IC~708 `N' and `S' components in Table 3 we have used 5 GHz as the break frequency.
For both IC~711 and IC~708, twin jets start from the compact head almost in the same way (Figures 6 and 7) and which conform to the 
basic model of jet-bending by Begelman et al. (1979), but in case of IC~711 they soon merge together to form a single tail while in 
case of IC~708, twin jets, though similar to those in  IC~711, go their separate ways (divorced!), forming individual lobes, and result in a WAT morphology. 
It is still not clear what really decides the 
formation of these very distinct morphologies. IC~708 as compared to  IC~711 is of course an order of magnitude smaller in size 
(unless geometric projections play a major role). Bliton et al (1998) have suggested that NATs are associated with dynamically complex clusters with possible recent or ongoing cluster-subcluster
mergers and the merger-induced bulk motion of the ICM causing the jet bending.
This may be supported, in some clusters, by the existence of NAT radio
galaxies with their tails oriented along the same direction (Feretti et al. 1999; Feretti et al. 2001; Feretti \& Giovannini 2008), since it seems unlikely that their parent galaxies are all moving
towards the same direction. But in Abell 1314, presence of a WAT (IC~708) close to a NAT (IC~711) 
may cause doubt on this hypothesis.  
From Coleman et al. (1976), the parent optical galaxies IC~708 and IC~711 do not seem to differ 
much in their observed radial speeds, which also indicates that their orientations with respect to the observer's line of sight 
and thus the geometric projection effects may not be very different from each other. It does not seem that the ICM density could 
have very large gradients to cause large difference at the positions of these two radio galaxies. This is because IC~711, with its very long tail, 
samples a fairly large fraction of the cluster expanse and the tail does not seem to vary much in its width over its large extent, and assuming that the diffuse parts of the radio tail are in pressure equilibrium with it, the pressure within the ICM could not be varying too much over its expanse.
However, estimates of $U_{\rm min}$ for IC~711 and IC~708 do differ by about a factor of three and the radio emission in IC~708 might be three to five times younger than that in IC~711 (Table 3). However, the difference in morphologies could not be just due to their different ages as one cannot visualize a WAT like IC~708 with time turning into a NAT like IC~711 or vice versa.

In addition to IC~711 and IC~708, another cluster member, IC~712, the galaxy near the centre of the cluster, may have a head-tail structure (Feretti \& Giovannini 1994). In our observations, it remains an unresolved radio source, however, there is some diffuse emission south-east of the source, which, assuming it is genuine, may or may not be associated with IC~712. The low frequency LOFAR map shows faint extended emission east of IC~712 (Wilber et al. 2019). The radio spectrum of radio galaxy IC~712 is flat below 610 MHz, but at higher frequencies it has a normal radio spectrum (Table 3).  Feretti \& Giovannini (1994) have claimed that it is the smallest head-tail radio galaxy known, with its radio structure ($\sim 6.5$ kpc for our adopted Hubble constant $H_{0}=70\,$km~s$^{-1}$\, Mpc$^{-1}$) lying well within the optical extent of the galaxy. If this is confirmed, then it might cast doubts on the  prevalent idea that the radio tail results from the cluster galaxy ploughing through the intra-cluster gas, as the radio tail structure of IC~712 is embedded well within the extent of the optical galaxy itself. Otherwise one may have to admit that the cluster environment influences the morphology of a radio source even well within the optical galaxy extent (Feretti \& Giovannini 1994).

\section{CONCLUSIONS}
We presented low-frequency, GMRT 
observations at 240, 610 and 1300 MHz of IC~711, a long head-tail radio galaxy, in fact the longest head-tail radio galaxies known. 
Our high resolution 610 MHz map showed a long straight tail of angular extent $\sim 10$ arcmin, and later two sharp bends in the tail. 
The radio structure, 
especially the long straight tail, does not seem to be consistent with a simple circular motion around the cluster centre, contrary to a previous suggestion in the literature. 
Each of the bends in the tail of IC~711 seems to occur very close to a prominent radio source, though it does not seem  that either of 
the bends is really related to the respective radio object. These two radio sources close to the bends are most likely unrelated to the cluster and are background objects. Another radio source `U1' in the cluster could  curiously be either a double source with an optical identification with a 20 mag galaxy within the radio emission region at 610 MHz, or as seen at 1300 MHz, it could be of a head-tail morphology but with no cluster galaxy coinciding with its flat spectrum head.
The spectral index of the tail of IC~711 steadily steepens as we move away from the head, with the end parts of the tail showing the steepest spectrum ($\alpha \, {\sim}$ 4 -- 5) ever seen in any diffuse radio emission region. Moreover, a break in the spectrum is seen all along the tail. 

\section*{ACKNOWLEDGEMENTS}
We thank an anonymous referee for his/her suggestions that helped improve the paper. We acknowledge Heinz Andernach for his comments on the manuscript. 
We thank the staff of the GMRT that made these observations possible. 
GMRT is run by the National Centre for Radio Astrophysics of the Tata Institute of Fundamental Research.
Funding for the Sloan Digital Sky Survey (SDSS) IV has been provided by the
Alfred P. Sloan Foundation, the U.S. Department of Energy Office of
Science, and the Participating Institutions. SDSS acknowledges
support and resources from the Center for High-Performance Computing at
the University of Utah. This research has made use of the NASA/IPAC extragalactic database (NED)
which is operated by the Jet Propulsion Laboratory, Caltech, under contract
with the National Aeronautics and Space Administration.
\begin{appendix}
\section{MINIMUM ENERGY CALCULATIONS IN THE CASE OF A SPECTRAL BREAK}
We derive here an expression for minimum energy in a synchrotron source where a break is seen in the power-law spectrum of the observed flux-density. 
Such a break 
can occur in the synchrotron spectrum due to radiation losses in the case where, e.g., there is an isotropic distribution of electrons or if  in the 
radiating region there is a continuous injection of fresh electrons with a power law energy distribution (Kardashev 1962; Pacholczyk 1970).  We use cgs 
system of units throughout.

A relativistic electron of rest mass $m_{0}$ and energy $E$ (corresponding to a Lorentz factor $\gamma$), gyrating in a uniform 
magnetic field $B$ with a gyro frequency $\nu_{\rm g}= \omega_g/2\pi=e\, B/(2\pi m_{0} c\,\gamma)$, 
emits most of its radiation in a narrow frequency band
around its characteristic synchrotron frequency (Rybicki \& Lightman 1979)
\begin{equation}
\label{eq:64b}
\nu_{\rm c} = \frac {3}{2}\,\nu_{\rm g} \sin\theta \;\gamma^{3}= 
\frac {3}{4 \pi}\frac{e\, B_\perp}{m_{0}\, c}\;\gamma^{2}=c_1 \, B_\perp E^{2}.
\end{equation}
where $c_1= 3e/(4\pi m_{0}^3\, c^5)=6.26 \times 10^{18}$ (Pacholczyk 1970)
and $\theta$ is the pitch angle. 
Accordingly, with the assumption that almost all the radiation from an electron is at a frequency 
$\nu \approx \nu_{\rm c}$, we can write a one-to-one relation between the frequency at which synchrotron 
radiation is being emitted and the energy $E$ of the electron emitting that radiation
\begin{equation}
\label{eq:64c}
E \approx \left({\frac {\nu}{c_1 B_\perp}}\right)^{1/2}.
\end{equation}
In radio sources, the
observed flux density in the optically thin part of the spectrum
usually follows a power law, i.e., $I_{\nu} \propto \nu^{-\alpha} $,
between the lower and upper cut off frequencies $\nu_{1}$ and
$\nu_{2}$. In synchrotron theory this spectrum results from a power
law energy distribution of radiating electrons $N(E)=N_0
E^{-\xi}$ within some range $E_{1}$ and $E_{2}$, with $\xi =
2\alpha+1$ and $E_{1}$, $E_{2}$ related to $\nu_{1}$, $\nu_{2}$ by
Equation~(\ref{eq:64c}). In the case of a spectral break at say, $\nu_0$, 
there may be spectral indices $\alpha_1$ between $\nu_{1}$ and $\nu_{0}$ and 
$\alpha_2$ between $\nu_{0}$ and $\nu_{2}$, corresponding to energy 
indices $\xi_1$ between $E_{1}$ and $E_{0}$ and $\xi_2$ between 
$E_{0}$ and $E_{2}$.

Now the energy distribution of radiating electrons cannot be written 
simply as $N(E)=N_0 E^{-\xi_1}$ between $E_{1}$ and $E_{0}$ and 
$N(E)=N_0 E^{-\xi_2}$  between $E_{0}$ and $E_{2}$, as it will give 
$N(E_0)=N_0 E_0^{-\xi_1}=N_0 E_0^{-\xi_2}$ forcing $\xi_1=\xi_2$. 
Instead we can write it as $N(E)=N_0 (E/E_0)^{-\xi_1}$ and 
$N(E)=N_0 (E/E_0)^{-\xi_2}$ for the two ranges with $N_0$ as the 
number density at $E_0$.

A highly relativistic electron ($\gamma \gg 1$) radiates power at a rate (Melrose 1971; Jackson 1975; Longair 2011)
\begin{equation}
\label{eq:64i}
d E/d t = c_2 \, B_\perp^2 (1-1/\gamma^2) E^2 \approx c_2 \, B_\perp^2 E^2, 
\end{equation}
where $c_2= 2e^4/(3m_{0}^4\, c^7)=2.37 \times 10^{-3}$ (Pacholczyk 1970). 

Then luminosity $L$ for such a synchrotron source is given by,
\begin{eqnarray}
\nonumber
L&=& V \phi\:N_0\: c_2 \, B_\perp^2 \left[\int_{E_{1}}^{E_{0}} {(E/E_0)^{-\xi_1}}E^{2}\: {\rm d}E\right. \\
\label{eq:64j}
&&\left.+ \int_{E_{0}}^{E_{2}}{(E/E_0)^{-\xi_2}}E^{2}\: {\rm d}E  \right]
\end{eqnarray}
where $V$ is the volume of the source and $\phi$ is the volume filling factor. 
\begin{eqnarray}
\nonumber
L&=&  V\phi\:N_0\: c_2 \, B_\perp^2 E_0^3\left[\frac {(E_1/E_0)^{-\xi_1+3} -1 }{\xi_1-3}\right.\\
\label{eq:64k}
&&\left.-\frac {(E_2/E_0)^{-\xi_2+3}  -1}{\xi_2-3}  \right].
\end{eqnarray}
Using Equation (\ref{eq:64c}), we can write 
\begin{eqnarray}
\nonumber
L&=&  \frac {V\phi\:N_0\: c_2\: \nu_0^{1.5}\, B_\perp^{0.5}}{2c_1^{1.5}} \left[\frac {(\nu_1/\nu_0)^{-\alpha_1+1} -1 }{\alpha_1-1}\right. \\
\label{eq:64l}
&&\left.-\frac {(\nu_2/\nu_0)^{-\alpha_2+1}  -1}{\alpha_2-1}  \right]
\end{eqnarray}

The flux-density can also be written as $S_0 (\nu/\nu_0)^{-\alpha_1}$ and 
$S_0 (\nu/\nu_0)^{-\alpha_2}$ for the two ranges $\nu_{1}$ -- $\nu_{0}$ and 
 $\nu_{0}$ -- $\nu_{2}$, with $S_0$ as the flux-density at $\nu_0$.
Then the luminosity of such a source can be evaluated from the observed flux density 
at the spectral break as 
\begin{eqnarray}
\nonumber
L&=&  4\pi D_L^2 S_0 \left[\int_{\nu_{1}}^{\nu_0} {(\nu/\nu_0)^{-\alpha_1}}\: {\rm d}\nu\right. \\
\label{eq:64d}
&&\left.+ \int_{\nu_0}^{\nu_2}(\nu/\nu_0)^{-\alpha_2}\: {\rm d}\nu  \right],
\end{eqnarray}
or
\begin{eqnarray}
\nonumber
L&=&  4\pi D_L^2 S_0 \: \nu_0 \left[\frac {(\nu_1/\nu_0)^{-\alpha_1+1} -1 }{\alpha_1-1}\right. \\
\label{eq:64e}
&&\left.-\frac {(\nu_2/\nu_0)^{-\alpha_2+1}  -1}{\alpha_2-1}  \right].
\end{eqnarray}

From Equations~(\ref{eq:64l}) and (\ref{eq:64e}) we can evaluate $N_0$ as
\begin{equation}
\label{eq:64e1}
N_0= \frac{8\pi \:c_1^{1.5} D_L^2 S_0}{c_2 V\phi\: B_\perp^{0.5} \: \nu_0^{0.5}}
\end{equation}

Energy density of the relativistic electrons in a synchrotron radio
source is given by 
\begin{eqnarray}
\nonumber
U_{\rm e}&=& N_0 \left[\int_{E_{1}}^{E_{0}} {(E/E_0)^{-\xi_1}}E\: {\rm d}E\right.\\
\label{eq:64f}
&&\left.+ \int_{E_{0}}^{E_{2}}{(E/E_0)^{-\xi_2}}E\: {\rm d}E  \right],
\end{eqnarray}
or
\begin{eqnarray}
\nonumber
U_{\rm e}&=& N_0 E_0^2\left[\frac {(E_1/E_0)^{-\xi_1+2} -1 }{\xi_1-2}\right.\\
\label{eq:64g}
&&\left.-\frac {(E_2/E_0)^{-\xi_2+2}  -1}{\xi_2-2}  \right].
\end{eqnarray}

Using Equation(\ref{eq:64c}), energy density of the relativistic electrons in a synchrotron radio
source is then given by 
\begin{eqnarray}
\nonumber
U_{\rm e}&=& \frac {N_0\: \nu_0} {2c_1 \:B_\perp}
\left[\frac {(\nu_1/\nu_0)^{-\alpha_1+0.5} -1 }{\alpha_1-0.5}\right.\\
\label{eq:64h}
&&\left.-\frac {(\nu_2/\nu_0)^{-\alpha_2+0.5}  -1}{\alpha_2-0.5}  \right],
\end{eqnarray}
the expression to be evaluated in `limit' if any of the $\alpha's=0.5$. 

Substituting $N_0$ from Equation~(\ref{eq:64e1}), we get
\begin{eqnarray}
\nonumber
U_{\rm e} &=& \frac  {4\pi D_L^2 S_0\: \nu_0^{0.5}\: c_1^{0.5} } {V\phi\:c_2 \, B_\perp^{1.5}}\left[\frac {(\nu_1/\nu_0)^{-\alpha_1+0.5} -1}{\alpha_1-0.5}\right.\\
\label{eq:64n}
&&\left.-\frac {(\nu_2/\nu_0)^{-\alpha_2+0.5}  -1}{\alpha_2-0.5} \right].
\end{eqnarray}
The magnetic field energy density is $U_{m}= B^{2}/8 \pi$.
Therefore total energy density of the source is $U_{\rm t}= (1+k) U_{\rm e}+U_{m} = a B^{-1.5} + b B^{2}$, where $k$ represents the factor 
for energy in heavier particles (baryons), which may range anything from 1 to 2000 or so. It should be noted that if the plasma comprises both electrons and positrons (a pair plasma), 
then k=0, since the observed flux density gets contributions from both electrons and positrons, and $N_0$ then, from Equation~(\ref{eq:64e1}), is already inclusive of both electrons and positrons whose contributions are indistinguishable in incoherent synchrotron cases 
(Singal 2012). The expression for energy density is similar to as in the
standard case of a single spectral index (Pacholczyk 1970; Moffet 1972), except that the co-efficient $a$ is  somewhat different. 

After minimizing the total energy with $B$, we get
\begin{eqnarray}
\nonumber
U_{\rm min}& \!\!=&\!\!3.0 \times 10^{-10} \left[ \frac {(1+k)(1+z)^{3.5} \: S_{0}\:  \nu_{0} ^{0.5}\: }  { \phi \: \psi_{x}  \psi_{y}\: s\: \sin^{1.5}\!\theta}\right.\\ 
\nonumber
&&\!\!\!\!\!\!\times\!\!\left.\left\{\frac {(\nu_1/\nu_0)^{-\alpha_1+0.5} -1 }{\alpha_1-0.5}
-\frac {(\nu_2/\nu_0)^{-\alpha_2+0.5} -1}{\alpha_2-0.5}  \right\}\right]^{4/7} \\
\label{eq:64o}
&& {\rm erg\: cm}^{-3}. 
\end{eqnarray}
Here $z$ is the redshift, $S_{0}$ (Jy) is the flux density at the break frequency 
$\nu_{0}$ (GHz), $\alpha_1$ and $\alpha_2$ are the spectral indices (defined as S$\propto\nu^{-\alpha}$) before and after the break, with $\nu_{1}$ 
and $\nu_{2}$ (GHz) as the lower and upper cut off
frequencies presumably for the observed radio spectrum, $ \psi_{x}$ and $ \psi_{y}$ (arcsec) represent the size of the source component along its major 
and minor radio axes, 
$s$ is the path length through the component along the line of sight in kpc and $\phi$ is the volume filling factor. 
It should be noted that the expression within curly brackets on the right hand side in Equation~(\ref{eq:64o}) is independent 
of redshift if $\nu_{1}, \nu_0, \nu_{2}$ are all in observer's frame, 
determined, for instance, from observations. 

The magnetic field value corresponding to the minimum energy is given by
\begin{eqnarray}
\label{eq:64q1}
B_{\rm min} =\left[\frac {24 \pi\: U_{\rm min}}{7}\right]^{0.5}=3.3\: U_{\rm min}^{0.5}\; G.
\end{eqnarray}
It should be noted that $B_{\rm min}$ is not the minimum value of the magnetic field in the source under consideration. In fact it yields nearly the 
equipartition value of the magnetic field. 

For a single spectral index case, where $S_{0}$  is the flux density at any frequency 
$\nu_{0}$ with $\nu_{1} < \nu_0 < \nu_{2}$, we put $\alpha_1=\alpha_2=\alpha$ in Equation~(\ref{eq:64o}) to get,
\begin{eqnarray}
\nonumber
U_{\rm min}& =& 3.0 \times 10^{-10} \left[ \frac {(1+k)(1+z)^{3.5} \: S_{0}\:  \nu_{0} ^{\alpha}\: }  {\phi\: \psi_{x} \psi_{y}\: s\: \sin^{1.5}\!\theta}\right. \\
\label{eq:64p}
&&\!\!\!\!\!\!\times\!\!\left.\left\{\frac {\nu_1^{-\alpha+0.5} - \nu_2^{-\alpha+0.5}}{\alpha-0.5} \right\}\right] 
^{4/7} {\rm erg\: cm}^{-3}. 
\end{eqnarray}
If $\nu_{1}$ and $\nu_{2}$ represent the cut-off frequency values in the rest frame of the source (specified based on some other 
arguments), then the factor $(1+z)^{3.5}$ in Equation~(\ref{eq:64p}) gets replaced with  $(1+z)^{3+\alpha}$ to yield the same expression as available in the 
literature (Miley 1980; Singal et al. 2004).
\section{THE PITCH ANGLE PARADOX AND RADIATIVE LIFE TIMES IN A SYNCHROTRON SOURCE}
Consider a charge particle in its gyrocentre (GC) frame ${\cal S}'$, where 
the charge has no component of velocity parallel to the magnetic field and has 
only a circular motion in a plane perpendicular to the magnetic field (with a pitch angle $\theta=\pi/2$). 
In this frame, due to radiative losses by the charge, there will be a decrease in the velocity which is solely  
in a plane perpendicular to the magnetic field, and the charge will never ever attain a velocity component parallel to the magnetic field because of radiation losses.

Now we look at this particle from the lab frame ${\cal S}$, in which the charge has (at least to begin with) a motion, $\beta_{\parallel}=v_{\parallel}/c$ along 
the magnetic field. Since in the inertial frame ${\cal S}'$, the charge never gets a velocity component parallel 
to the magnetic field  and the two inertial frames (${\cal S}$ and ${\cal S}'$) continue to move with reference to each other with a 
constant $\beta_\parallel$, the parallel component of velocity of the charge should remain unchanged even in ${\cal S}$. However, magnitude 
of the perpendicular component of velocity is continuously decreasing because of radiative losses, therefore the pitch angle of the charge, 
$\theta = \tan^{-1} (\beta_\perp/\beta_\parallel)$, should decrease continuously with time and 
the velocity vector of the charge should increasingly align with the magnetic field vector.
This clearly shows the inadequacy of the standard picture of radiation losses which had led to the conclusion that 
the pitch angle of the charge is a constant; arguments based on the relativistic transformations have shown us that for a charge undergoing synchrotron losses, the pitch angle would be progressively reducing with time (See Singal 2018 for a more comprehensive discussion on this issue).

In lab frame ${\cal S}$, both $\sin \theta$ and $\gamma$ are time-dependent variables in Equation~(\ref{eq:30.2}), making the solution of the equation to be a cumbersome process. However, in GC frame ${\cal S}'$ the pitch angle remains always a constant ($\theta'=\pi/2$). Then we have 
\begin{equation}
\label{eq:30.7}
\frac{{\rm d}{\gamma}'}{{\rm d}t'} =  -{\eta}\beta'^2\gamma'^2= -{\eta}(\gamma'^2-1)\;.
\end{equation}
Now Equation~(\ref{eq:30.7}) has a solution
\begin{equation}
\label{eq:30.2c1}
\tanh^{-1}\frac{1}{\gamma'}= {\eta} t' + a\;.
\end{equation}
Let ${\gamma'_{\rm o}}$ be the initial energy at $t'=0$ in frame ${\cal S}'$, then 
$1/{\gamma'_{\rm o}}=\tanh(a)$ and at time $t'=\tau'$ we have
\begin{equation}
\label{eq:30.2e3}
\tanh^{-1}\frac{1}{\gamma'} = \tanh^{-1}\frac{1}{{\gamma'_{\rm o}}}+{\eta} \tau'\;.
\end{equation}
which complies with the expectations that as $\tau' \rightarrow \infty$, $\gamma' \rightarrow 1$.

A transformation between ${\cal S}'$ and ${\cal S}$ gives 
$\gamma' \gamma_\parallel  = \gamma$ and  $\gamma' \beta' = \gamma \beta_\perp$ or $\beta' = \gamma_\parallel \beta_\perp$ (Melrose 1971), 
where $1/\gamma_\parallel=\surd(1- \beta_\parallel^2)$. 
Also we have ${\rm d}t/{\rm d}t'=\gamma_\parallel$ or $\tau = \tau' \gamma_\parallel$. 
For the transformation of acceleration we then get $\dot{\beta}'  = \gamma_\parallel^2 \dot{\beta}_\perp$  
with $\dot{\beta}_\parallel = \dot{\beta}'_\parallel/\gamma_\parallel^3 = 0$. 

Equation~(\ref{eq:30.2e3}) can be transformed in terms of quantities expresses in the lab frame ${\cal S}$
\begin{equation}
\label{eq:30.2e4}
\tanh^{-1}\frac{\gamma_\parallel}{\gamma} = \tanh^{-1}\frac{\gamma_\parallel}{{\gamma_{\rm o}}}+\frac{{\eta} \tau}{\gamma_\parallel}\;.
\end{equation}
Thus we get $\gamma$ as a function of time, starting from an initial $\gamma_{\rm o}$. 

Equation~(\ref{eq:30.7}) can be also written as
\begin{equation}
\label{eq:30.8a}
\gamma'^3 \dot{\beta}' {\beta'} =  -{\eta} \beta'^2 {\gamma}'^2,
\end{equation}
It should be noted that magnetic field does no work and any change in $\gamma'$ is only due to the radiation losses and $\dot{\beta}'$ is in a direction 
opposite to $\beta'$ in the GC frame. 

We simplify Equation~(\ref{eq:30.8a}) to write 
\begin{equation}
\label{eq:30.8b}
\frac{\dot{\beta}'}{\beta'} =  \frac{-{\eta}} {\gamma'}\;.
\end{equation}

Then transforming to the lab frame ${\cal S}$ we have
\begin{equation}
\label{eq:30.8c}
\frac{\dot{\beta}_\perp}{\beta_\perp} =  \frac{-{\eta}} {\gamma} \;.
\end{equation}
Both $\beta$ and $\theta$ appearing in $\beta_\perp=\beta \sin \theta$ are functions of time. Also $\dot{\beta}_\parallel={\rm d}(\beta \cos \theta)/{\rm d}t=0$. 
Thence we get
\begin{equation}
\label{eq:30.6e}
\frac{{\rm d}\theta}{{\rm d}t}= \frac{-{\eta}\sin 2\theta}{2\gamma}\;,
\end{equation}
which gives us the rate of change of pitch angle with time.

The expressions for the pitch angle changes and the radiation losses could be derived in an alternate way directly from the radiation reaction force 
(Singal 2016), where it is also shown that $1/\eta$ is the characteristic decay time of a synchrotron electron over 
which it not only turns from ultra relativistic into mildly relativistic one, but due to the  change in the pitch angle during this time, its velocity 
vector also moves outside the angle $1/\gamma_{\rm o}$ around the line of sight towards the observer.

However, on a much shorter time scale $\tau \sim 1/(\gamma_{\rm o}\eta)$ where $\gamma_{\rm o}\gg 1$ and for not too small a 
pitch angle, $\sin \theta_{\rm o} \gg 1/\gamma_{\rm o}$, implying $\gamma_\parallel / \gamma_{\rm o} \ll 1$ 
(as $1/\gamma_\parallel=\surd(1-\beta_{\rm o}^{2}\cos^2 \theta_{\rm o})\approx \sin\theta_{\rm o}$), from Equation~(\ref{eq:30.2e4}) we could write
\begin{equation}
\label{eq:30.2f}
\frac{1}{\gamma} = \frac{1}{{\gamma_{\rm o}}}+\frac{{\eta} \tau}{\gamma_\parallel^2}\;,
\end{equation}
or
\begin{equation}
\label{eq:30.2g}
\gamma = \frac{\gamma_{\rm o}}{1+{\eta}\tau \gamma_{\rm o} / \gamma^2_\parallel}
=\frac{{\gamma_{\rm o}}}{1+{\eta}\tau\gamma_{\rm o}\sin^2 \theta_{\rm o}}\;.
\end{equation}
Equation~(\ref{eq:30.2g}) is the same relation as Equation~(\ref{eq:30.4}), but which now we know is valid only for 
$\gamma_{\rm o} \gg 1$, $\tau \stackrel{<}{_{\sim}}1/(\gamma_{\rm o}\eta)$ and for pitch angles not too small 
($\sin \theta_{\rm o} \gg 1/\gamma_{\rm o}$).

From Equations~(\ref{eq:30.2f}) and (\ref{eq:30.2g}) it is easily seen that irrespective of the initial energy $\gamma_{\rm o}$ of the radiating electron 
at time 
$t=0$, at a later time $t=\tau$, its energy will have an upper limit 
$\gamma^* = 1/\eta\tau\sin^2 \theta_{\rm o}$. This in turn tells us about the radiating life-times of the synchrotron electrons 
\begin{equation}
\label{eq:30.2h}
\tau_{syn}\approx \frac{16}{B_\perp^{2}\gamma^*} \;\;{\rm yr}.
\end{equation}
From Equation~(\ref{eq:64b}), a synchrotron electron with energy $\gamma$ radiates around
\begin{equation}
\label{eq:30.2h1}
\nu \approx{4.20 \times 10^6}\,B_\perp \gamma^{2},
\end{equation}
with an exponential drop in flux density above frequency $\nu$. Therefore an upper limit $\gamma^*$ implies an upper limit on frequency $\nu^*$ above which 
the flux density falls exponentially. 

From a sharp steepening of the observed spectrum one gets the value of $\nu^*$  and then an estimate the age of the synchrotron source is obtained as
\begin{eqnarray}
\nonumber
\frac{\tau_{syn}}{\rm yr}& = &\frac{3.35 \times 10^4}{B_\perp^{1.5} \sqrt{\nu^*}}\\
\label{eq:30.2i}&= &1.06 \times 10^9\left(\frac{B_\perp}{\mu{\rm G}}\right)^{-1.5} 
\left(\frac{\nu^*}{\rm GHz}\right)^{-0.5},
\end{eqnarray}
provided an estimate of magnetic field $B$ is available, e.g., from the minimum energy estimates (Equation~(\ref{eq:64q1})) which yields nearly an 
equipartition value of the  magnetic field.
\end{appendix}\\
{}
\end{document}